\documentstyle[aps,prd,eqsecnum,amssymb,floats,epsfig]{revtex}

\newcommand{\be}{\begin{equation}}
\newcommand{\ee}{\end{equation}}
\newcommand{\bea}{\begin{eqnarray}}
\newcommand{\eea}{\end{eqnarray}}

\newcommand{\dst}{\displaystyle}

\def\np{({\bf n}\cdot{\bf p})}
\def\pp{{\bf p}^2}
\def\ppp{({\bf p}^2)}

\def\omk{\omega_{\text{kinetic}}}
\def\oms{\omega_{\text{static}}}

\begin{document}

\title{
On the determination of the last stable orbit
for circular general relativistic binaries
at the third post-Newtonian approximation}

\author{Thibault Damour}
\address{Institut des Hautes \'Etudes Scientifiques,
91440 Bures-sur-Yvette, France}

\author{Piotr Jaranowski}
\address{Institute of Theoretical Physics,
University of Bia{\l}ystok,
Lipowa 41, 15-424 Bia{\l}ystok, Poland}

\author{Gerhard Sch\"afer}
\address{Theoretisch-Physikalisches Institut,
Friedrich-Schiller-Universit\"at,
Max-Wien-Platz 1, 07743 Jena, Germany}

\maketitle

\begin{abstract}
We discuss the analytical determination of the location of the Last Stable Orbit
(LSO) in circular general relativistic orbits of two point masses.  We use
several different ``resummation methods'' (including new ones) based on the
consideration of gauge-invariant functions, and compare the results they give at
the third post-Newtonian (3PN) approximation of general relativity.  Our
treatment is based on the 3PN Hamiltonian of Jaranowski and Sch\"afer.  One of
the new methods we introduce is based on the consideration of the (invariant)
function linking the angular momentum and the angular frequency.  We also
generalize the ``effective one-body'' approach of Buonanno and Damour by
introducing a non-minimal (i.e.\ ``non-geodesic'') effective dynamics at the 3PN
level.  We find that the location of the LSO sensitively depends on the
(currently unknown) value of the dimensionless quantity $\oms$ which
parametrizes a certain regularization ambiguity of the 3PN dynamics.  We find,
however, that all the analytical methods we use numerically agree between
themselves if the value of this parameter is $\oms\simeq-9$. This suggests
that the correct value of $\oms$ is near $-9$ (the precise value
$\oms^*\equiv-\frac{47}{3}+\frac{41}{64}\pi^2=-9.3439\ldots$ seems to play a 
special role).  If this is the case, we then show how to further improve the 
analytical determination of various LSO quantities by using a ``Shanks'' 
transformation to accelerate the convergence of the successive (already 
resummed) PN estimates.
\end{abstract}


\section{Introduction}

The study of the late dynamical evolution of binaries made of compact objects
(neutron stars or black holes) is important because such systems are the most
promising candidate sources for interferometric gravitational-wave detectors
such as LIGO and VIRGO.  In particular, the global structure of the
gravitational waveform emitted by such a binary sensitively depends on the
frequency at which the system's orbital evolution changes from a
gravitational-radiation-driven inspiral phase to a plunge phase followed by
coalescence \cite{DIS,DIS2}.

In the test-mass limit $(\mu\ll M)$ the orbital dynamics is that of a test
particle (of mass $\mu$) moving in a Schwarzschild background (of mass $M$).  A
very important qualitative feature of circular orbits in such a background is
the existence of a Last Stable Orbit (LSO) located at the (area) radius
$R_{\text{LSO}}=6GM/c^2$.  When considering the effect of
gravitational-radiation reaction, one expects the test-particle motion to change
abruptly near $R_{\rm LSO}$ from a slow inspiral to a fast plunge.  By analogy,
one believes that the motion of a compact binary made of comparable masses,
$m_1$ and $m_2$, will exhibit a similar transition from inspiral to plunge, with
the location of the transition being mainly determined by the existence of a
Last Stable Orbit in the {\it conservative} part of the two-body Hamiltonian.

Several authors have tried to estimate the location of the LSO in
comparable-mass (compact) binaries.  An early analytical estimate was made by
Clark and Eardley \cite{CE77} using the first post-Newtonian (1PN) Hamiltonian.
Some authors \cite{BD92,C94,B00} tried to use initial value formalisms to locate
the LSO.  However, the initial value approaches used in these works assume a
conformally flat metric, and therefore do not correctly incorporate the well
known second post-Newtonian (2PN) dynamics \cite{DD,DS85,DS88}.  In this work we
shall take the view that a correct incorporation of all 2PN effects is a
necessary (if maybe not sufficient) prerequisite for an accurate determination
of the location of the LSO.  Previous treatments that used the full 2PN dynamics
to try to analytically determine the location of the LSO include
Refs.\cite{KWW,WS93,SW93,DIS,BD99}.  The aim of the present work is to extend
these PN-based analytic determinations of the LSO to the third post-Newtonian
(3PN) level.  The conservative part of the 3PN Hamiltonian for two point masses
has been obtained in 1998 by Jaranowski and Sch\"afer \cite{JS98}, though with
some remaining ambiguity due to the need to regularize the divergent integrals
entailed by the use of point-like sources.  The determination of the 3PN
dynamics has been recently completed by deriving the Hamiltonian in a
non-mass-centered frame, and by fixing a certain momentum-dependent
regularization ambiguity \cite{DJS2}.  Recently, we have extracted from the 3PN
Hamiltonian of Ref.\ \cite{JS98} all its {\it dynamical invariants}, i.e.\ all
the functions linking dynamical quantities which do not depend on the choice of
coordinates in phase-space \cite{DJS1}.  In this paper, we shall use several of
these 3PN invariants to determine the location of the LSO.  Some of the methods
of LSO determination that we shall use below generalize previous works
\cite{DIS,BD99}, but others are new.

\section{Methods directly based on dynamical invariants}

Before embarking on the discussion of the methods we shall use here to extract
the LSO from PN expansions, let us stress that the basic theme underlying our
endeavours is the following:  Our problem is to extract some
semi-non-perturbative information from (badly convergent) perturbation
expansions.  We shall do that by using several types of ``resummation methods''.
The basic idea of all resummation methods is simply the following: to complete
the information contained in the first few terms of a perturbative expansion
$f(z)=c_0+c_1\,z+\cdots+c_n\,z^n+{\cal O}(z^{n+1})$ by injecting some
non-perturbative information about the global behaviour (if possible in the
complex plane) of the exact function $f(z)$.  The amount of global information
one has about the function $f(z)$ determines the best type of resummation 
method to use.  For instance, if
the only information at our disposal is that the function $f(z)$ is (probably)
meromorphic in the complex $z$-plane, then the best, all-purpose method is to
use Pad\'e approximants.  If we knew more about the location of the
singularities of $f(z)$ in the complex plane one might contemplate to use other
methods (e.g.\ change of independent variable, Borel transform, $\ldots$).

In this paper, we shall use two distinct classes of methods
for extracting the (invariant) location of the LSO from
the knowledge of the PN-expanded dynamics. The first class of methods
(which was introduced in Ref.\ \cite{DIS}) is discussed in the present Section.
The second class will be discussed in the next Section.
The first class of methods is based on the combination of three ideas:

(i) to work only with invariant functions;

(ii) to make a maximal use of the known, exact functional form of invariant 
functions in the test-mass limit $(\nu\equiv m_1m_2/(m_1+m_2)^2\rightarrow0$) 
and to assume some structural stability when the parameter $\nu$ is turned on;

(iii) to use Pad\'e approximants to represent the invariant functions which are 
(because of (ii)) expected to be meromorphic functions of their argument.

This method was applied in Ref.\ \cite{DIS} to the two invariant functions which 
play an essential role in the gravitational-damping-driven inspiral of a binary 
system: the binding energy $E(x)$ of a circular orbit, and the 
gravitational-wave flux $F(x)$ emitted by a circular orbit, both being 
considered as functions of the (dimensionless) invariant parameter
\be
\label{eq2.1}
x \equiv \left(\frac{GM\omega}{c^3}\right)^{2/3}\,,
\ee
where $M\equiv m_1+m_2$ denotes the total mass of the binary, and $\omega$ the 
orbital angular frequency along a circular orbit.

Before introducing some variations on this method, let us motivate the 
interest of using the three ideas (i)--(iii) above. First, we recall that 
the PN expansions of non-invariant, i.e.\ gauge-dependent, functions can 
have (and do have, in some gauges) worse convergence properties than the 
PN expansions of invariant functions. For instance, the PN expansion of 
the gravitational-wave flux, from a test mass in circular orbit, say 
$F_{\rm TM}$, considered as a function of the harmonic-coordinate 
parameter $\gamma\equiv GM/(c^2 r_{\rm harmonic})$, is such that 
the 1PN ``correction'' to the leading ``quadrupole'' result becomes 
fractionally larger than 100\% (while being negative!) for a radius 
$r_{\rm harmonic}$ larger than the LSO (which is at $r_{\rm LSO}=5GM/c^2$ in 
harmonic coordinates). More precisely, if one formally writes down the expansion 
of $F_{\rm TM}(\gamma)$, in powers of $\gamma$, near the LSO (i.e.\ for $\gamma$ 
near $1/5$) one gets a series numerically proportional to:
$1 - 1.74\,(5\gamma) + 1.12\,(5\gamma)^{3/2} + 1.29\,(5\gamma)^2 + \cdots$, 
i.e.\ a series whose first terms do not exhibit any convergence near the LSO. By 
contrast, the flux $F_{\rm TM}$ expanded in terms of the invariant parameter 
$x$, Eq.\ (\ref{eq2.1}), ($x_{\rm LSO} = 1/6$ in the test-mass limit) has a 
more reasonable expansion proportional to
$1 - 0.619\,(6x) + 0.855\,(6x)^{3/2} - 0.137\,(6x)^2 + \cdots$. Still, it is 
clear that one needs some resummation technique for summing a series as slowly 
converging as $F_{\rm TM}(x)$. It was shown in great detail in Ref.\ 
\cite{DIS}, by making use both of the known analytical results on high-order 
(5.5PN) terms in the post-Newtonian expansion of the test-mass flux function 
$F_{\rm TM}(x)$ \cite{TTS96}, and of the existence of a pole-like blow up of 
$F_{\rm TM}(x)$ at $x=x_{\text{light ring}}=1/3$, that one could considerably 
speed up the convergence of the straightforward (Taylor-like) PN series, by 
replacing it by a suitably defined sequence of Pad\'e approximants (see, 
notably, Fig.\ 3 in \cite{DIS}).

To further illustrate the idea (ii), and the need for an acceleration of 
convergence, let us recall the treatment of the energy function $E(x)$ 
introduced in \cite{DIS}. In this paper we follow \cite{DJS1} in defining the 
dimensionless energy function
\be
\label{eq2.2}
E \equiv \frac{{\cal E}^R-Mc^2}{\mu c^2}\,,
\ee
where ${\cal E}^R$ is the total (``relativistic'') energy of the binary system 
(including the rest-mass contribution), and where
\be
\label{eq2.3}
M \equiv m_1+m_2, \quad
\mu \equiv \frac{m_1 m_2}{m_1+m_2}, \quad
\nu \equiv \frac{\mu}{M} = \frac{m_1 m_2}{(m_1+m_2)^2} \,.
\ee
For notational simplicity we shall henceforth use units such that $c=1$.
Note that the energy function used in \cite{DIS} was
$E^{\rm DIS}\equiv({\cal E}^R-M)/M=\nu E^{\rm here}$.

Because the argument $x$ is, from its definition (\ref{eq2.1}), of formal order 
${\cal O}(c^{-2})$, the knowledge of the dynamics at, say, the $n$th 
post-Newtonian ($n$PN) order entails the knowledge of the expansion of the ratio 
$E(x)/x$ up to $x^n$:
\be
\label{eq2.4}
E(x;\nu) = -\frac{1}{2}\,x \left[ 1 + E_1(\nu) \, x + E_2(\nu) \, x^2 + \cdots 
+ E_n(\nu) \, x^n + {\cal O}(x^{n+1}) \right] \,. 
\ee
[The term $-\frac{1}{2}\,x$ corresponds to the Newtonian binding energy
${\cal E}^{\rm NR}=-\frac{1}{2}\,\mu\,v^2$. Then the term $E_1(\nu)\,x$, for 
instance, represents the fractional 1PN effects, etc.] 
The symmetric mass ratio $\nu$, Eq.\ (\ref{eq2.3}), enters the expansion 
coefficients $E_n(\nu)$ as a parameter. At present, only $E_1 (\nu)$, 
$E_2(\nu)$ and $E_3(\nu)$ are known (with some ambiguity for the 3PN 
coefficient $E_3(\nu)$). They were written down in \cite{DJS1} and will 
be repeated below. On the other hand, in the test-mass limit $\nu 
\rightarrow 0$, the coefficients $E_n (0)$ are known (in principle) for 
any $n$. What is more, the exact expression of $E(x;\nu=0)$ is known:
\be
\label{eq2.5}
E(x;\nu=0) = \frac{1-2x}{\sqrt{1-3x}} - 1 \,.
\ee

It is known (see, e.g., \cite{DIS}) that the location of the LSO corresponds 
just to the minimum of the function $E(x)$ ($(dE(x)/dx)_{x_{\text{LSO}}}=0$). 
Therefore, it would seem that the most straightforward way of locating the LSO 
is to consider the successive ``Taylor approximants'' of $E(x)$, say 
$E_{T_n}(x)$ (defined for each integer $n$ as the R.H.S.\ of Eq.\ (\ref{eq2.4}), 
without the ${\cal O}\,(x^{n+1})$ error term), and to solve the equations 
$dE_{T_n}(x)/dx=0$. Let us see what this gives in the test-mass limit where the 
Taylor expansion of Eq.\ (\ref{eq2.5}) is known:
\be
\label{eq2.6}
E(x;\nu=0) = - \frac{1}{2} \, x \left( 1 - \frac{3}{4} \, x - 
\frac{27}{8} \, x^2 - \frac{675}{64} \, x^3 - \frac{3969}{128} \, x^4 - 
\cdots \right) \,.
\ee
The successive ``Taylor'' estimates of $x_{\rm LSO}$, or better
$x_{\rm LSO}^{T_n} / x_{\rm LSO}^{\rm exact}\equiv6\,x_{\rm LSO}^{T_n}$, are 
found to be (in the test-mass case):
\be
\label{eq2.7}
6\,x^{T_1} = 4\,; \quad
6\,x^{T_2} = 1.49284\,; \quad
6\,x^{T_3} = 1.17565\,; \quad
6\,x^{T_4} = 1.07680\,.
\ee
[To avoid confusion note that we use here the convention that `$T_n$' 
corresponds to the $n$PN approximation, i.e.\ a $(v/c)^{2n}$-accurate result, 
while in Ref.\ \cite{DIS} `$T_n$' referred to $(v/c)^{n}$-accuracy, i.e.\ to the 
$\frac{n}{2}$PN approximation.]
{From} Eq.\ (\ref{eq2.1}) the corresponding values of the orbital frequency 
at the LSO, scaled to the exact value,
$\widehat{\omega}\equiv\omega/\omega^{\rm exact}=(6x)^{3/2}$, read
\be
\label{eq2.8}
\widehat{\omega}^{T_1} = 8\,; \quad
\widehat{\omega}^{T_2} = 1.82398\,; \quad
\widehat{\omega}^{T_3} = 1.27472\,; \quad
\widehat{\omega}^{T_4} = 1.11739\,.
\ee

As the value of the frequency at the LSO is the most important 
observable one wishes to know (for data analysis purposes), one should 
reject any method which does not have the prospect of determining it to, 
say, better than about 10\%. The test-mass results\footnote{The logic 
here is to use the known convergence properties of the $\nu = 0$ limit 
to estimate the convergence when $\nu \ne 0$. It is, indeed, unlikely 
that turning on $\nu$ will drastically {\em improve} the convergence 
properties.} (\ref{eq2.8}) suggest that, if the dynamics is known only 
up to the 3PN level, straightforward Taylor approximants of the energy 
function $E(x)$ do not converge fast enough to determine satisfactorily 
the location of the LSO. [The 4PN level might barely suffice, but it 
seems anyway excluded that one will be able to analytically derive the 
4PN dynamics.]

This preliminary discussion motivates the necessity of boosting up the
convergence of the series (\ref{eq2.4}) or (\ref{eq2.6}).  This is here that the
ideas (ii) and (iii) enter.  The exact test-mass result (\ref{eq2.5}) suggests
(under the assumption that the $\nu \ne 0$ case represents a structurally stable
deformation of the $\nu=0$ limit) that the function $E(x;\nu)$, for $\nu\ne0$,
will have a branch-cut singularity (at some point
$x_0=\frac{1}{3}+{\cal O}(\nu)$) in the complex $x$ plane.  It is, a priori, 
much better to work with functions which are {\em meromorphic} in the complex
plane, because we can then make use of Pad\'e approximants, which are efficient
tools for accurately representing meromorphic functions.  This suggests to work
with some (to be defined) new invariant energy function that is known to be
meromorphic in the test-mass limit.

Looking at Eq.\ (\ref{eq2.5}) one is tempted to consider the square of the 
function $1+E$. However, Ref.\ \cite{DIS} remarked that the usual test-mass 
limit definition of this function, namely ($m_2 \ll m_1$ being the test-mass 
orbiting $m_1$)
\be
\label{eq2.9a}
1+E \equiv \frac{{\cal E}_2^R}{m_2}
\ee
with
\be
\label{eq2.9b}
{\cal E}_{\rm tot}^R = m_1 + {\cal E}_2^R + {\cal O} (m_2^2)
= m_1 - \frac{p_1^{\mu} \, p_{2\mu}}{m_1} + {\cal O} (m_2^2) \,,
\ee
looks unnaturally asymmetric with respect to the labels 1 and 2. [Essentially, 
this asymmetry comes from the fact, hidden in the symmetric definition 
(\ref{eq2.2}), that ${\cal E}_2^R$ represents the sum of the rest-mass energy of 
$m_2$ alone and of the binding energy of the binary system.] Ref.\ \cite{DIS} 
therefore suggested to introduce the more symmetric function
\be
\label{eq2.10}
\varphi(s) \equiv \frac{s - m_1^2 - m_2^2}{2 \, m_1 \, m_2} \equiv 
\frac{({\cal E}^R)^2 - m_1^2 - m_2^2}{2 \, m_1 \, m_2}
\ee
of the Mandelstam invariant
$s\equiv-(p_1^{\mu}+p_2^{\mu})^2\equiv({\cal E}^R)^2$. Indeed, in the test-mass 
limit
\be
\label{eq2.11}
\varphi(s) \equiv - \frac{p_1^{\mu} \, p_{2\mu}}{m_1 \, m_2}
\simeq \frac{{\cal E}_2^R}{m_2} = 1 + E = \frac{1-2x}{\sqrt{1-3x}} \, . 
\ee
It is then natural to define the function $e(x)$ (or rather $1+e(x)$), by 
setting
\be
\label{eq2.12}
1 + e(x) \equiv \biglb(\varphi (s(x))\bigrb)^2
= \Bigg( \frac{({\cal E}^R)^2 - m_1^2 - m_2^2}{2 \, m_1 \, m_2} \Bigg)^2 \, .
\ee
Then, $1+e(x)$, being equal to $(1-2x)^2/(1-3x)$ in the test-mass limit, i.e.\ 
(after subtraction of the trivial constant 1)
\be
\label{eq2.13}
e(x;\nu=0) = -x\,\frac{1-4x}{1-3x}\,,
\ee
one expects the function $e(x;\nu)$ to be meromorphic in $x$ when $\nu\ne0$ 
(with a pole located at some $x_{\rm pole}=\frac{1}{3}+{\cal O}(\nu)$).

This finally leads to the following ``$P$-approximant''-improved method for 
locating the LSO: starting from the Taylor approximants ($T$-approximants) of 
the original $E(x)$ function, Eq.\ (\ref{eq2.4}), compute first the 
corresponding $T$-approximants of the new $e(x)$ function, say
\be
\label{eq2.14}
e(x) = -x \, \left[ 1 + e_1 (\nu) \, x + e_2 (\nu) \, x^2 + e_3 
(\nu) \, x^3 + {\cal O} (x^4) \right] \, .
\ee
Then construct a sequence of Pad\'es of the Taylor-expanded $e(x)$:
\be
\label{eq2.15}
e_{P_n}(x) \equiv P_{\ell}^k \left[ T_n \, [e(x)] \right] \,,
\ee
where $k+\ell=n$.  Here $k$ and $\ell$ are the degrees of the polynomials
$N_k(x)$ and $D_{\ell}(x)$ entering the Pad\'e $P_{\ell}^k(x)=N_k
(x)/D_{\ell}(x)$.  It is known that, generically, the Pad\'e improvements are
best when one is near the ``diagonal'', i.e.\ when $\vert{k-\ell}\vert$ is
``small'' compared to $k$ and $\ell$.  When dealing with a function $f(x)$ that
is expected to have a {\em pole} at some $x_0\ne0$, one imposes the constraint
$\ell > 0$.  At 1PN order this uniquely fixes the values of $k$ and $\ell$,
namely $k = 0$ and $\ell = 1$.  At 2PN order the Pad\'e closest to the diagonal
is that with $k = 1$ and $\ell = 1$.  At 3PN order there are two possible
Pad\'es near the diagonal, namely $k=1$ and $\ell=2$, or $k=2$ and $\ell=1$.  In
this work we shall use the latter one ($ P^2_1$) because we found it to be more
robust under variations of the coefficients of the Taylor expansion of the
Pad\'eed function. (One aspect of this robustness is that the existence of a
real pole in this Pad\'e is always ensured, while this is not the case for the
other 3PN possibility:  $P^1_2$.)

Note also that Pad\'es are originally defined only for series of the 
regular type $\sigma(x)=c_0+c_1\,x+\cdots$ with $c_0\ne0$. When 
dealing with a function of the type $f_p(x)=x^p\,\sigma(x)$ with 
some relative integer $p$, we shall, {\em by convention}, {\em define} any 
Pad\'e of $f_p(x)$ as being $(k+\ell=n)$
\be
\label{eq2.16}
P_{\ell}^k [T_n[f_p (x)]] \equiv x^p \, P_{\ell}^k [T_n[x^{-p}\,f_p(x)]] \, . 
\ee

The ``$e_{P_n}$-estimate'' of the location of the LSO is then defined as 
the value $x_{P_n}$ at which $e_{P_n} (x)$ reaches a minimum. [It is 
easily seen that $e(x)$ follows the variations of $E(x)$, and in 
particular that it reaches a minimum at the same place as $E(x)$.]

We spent some time explaining in detail on one example our general 
Pad\'e-improved--test-mass-limit-motivated approach because we are going 
to extend it to several other invariant functions. In this (and the 
next) section, we present our various methodologies. The 3PN results 
obtained by them will be presented in a later section.

The introduction of the function $e(x)$ has two defects. First, it is 
not unique because we do not know for sure whether the function 
$(\varphi (s))^2$, Eq.\ (\ref{eq2.12}), is a ``better'' invariant than 
$(1+E)^2 = (1 + (\sqrt s - M) / \mu)^2$. Second, the Padeing of $e(x)$ 
starts giving meaningful results only at the 2PN level. Indeed the 1PN 
expansion of $e(x)$, in the test-mass limit, $e (x; \nu = 0) = -x \, 
(1-x + {\cal O} (x^2))$, yields a Pad\'e $e_{P_1} (x; \nu = 0) = -x / 
(1+x)$ which contains no pole on the positive real axis, and which 
formally predicts an LSO (minimum of $e_{P_1}(x)$) located at $x=+\infty$.

In this paper, we propose to consider another invariant function, which 
is more uniquely defined, and which gives sensible results already at 
the 1PN level. Let us consider the reduced angular momentum
\be
\label{eq2.17}
j \equiv \frac{\cal J}{\mu GM} = \frac{\cal J}{G m_1 m_2} \,,
\ee
where ${\cal J}$ denotes the total angular momentum of the system. In 
the test-mass limit the invariant function giving the (dimensionless) 
quantity $j$ in terms of the quantity $x$, Eq.\ (\ref{eq2.1}), reads
\be
\label{eq2.18}
j \, (x;\nu = 0) = \frac{1}{\sqrt{x (1-3x)}} \, .
\ee

This motivates the consideration of the squared (reduced) angular 
momentum $j^2(x)$ which is expected, when $\nu \ne 0$, to be a 
meromorphic function of $x$, with a pole at the ``light ring'' $x_{\rm 
pole} = \frac{1}{3} + {\cal O}(\nu)$. Therefore we propose to work with 
the Padeed form of $j^2$:
\be
\label{eq2.19}
j_{P_n}^2 (x;\nu) \equiv P_{\ell}^k \, [T_n [j^2 (x;\nu)]] \, ,
\ee
with $k+\ell=n$, and the choice of $\ell>0$ discussed above.
(We use in Eq.\ (\ref{eq2.19}) the convention (\ref{eq2.16}), i.e.\ the factor 
$x^{-1}$ in $j^2(x)$ is factored before taking a Pad\'e.)
Note that, in the test-mass limit, if we knew only the 1PN approximation 
to the function $j^2(x)$, i.e.\
$j^2(x;\nu=0)=x^{-1}\biglb(1+3x+{\cal O}(x^2)\bigrb)$, the procedure 
(\ref{eq2.19}) would reconstruct the {\it exact} result:
$j_{P_1}^2(x;\nu=0)=P_1^0[T_1[j^2(x;\nu=0)]]=[x(1-3x)]^{-1}$.

It is important to note that the perturbative information contained in 
the PN expansion of the function $j(x)$ (or equivalently $j^2(x)$) is 
totally equivalent (at any PN accuracy) to the information contained in, 
either the original energy function $E(x)$, or the new one $e(x)$, 
Eq.\ (\ref{eq2.12}). Indeed, the generic Hamiltonian equation 
$\dot{\theta}_i=\omega_i=\partial{H}/\partial{I_i}$ in action-angle 
variables $(I_i,\theta_i)$ yields
\be
\label{eq2.20}
\omega_{\rm circular} = \frac{d{\cal E}^R}{d{\cal J}}
= \frac{1}{GM} \frac{dE}{dj} \,,
\ee
which implies the identity
\be
\label{eq2.21}
\frac{dE(x)}{dx} = x^{3/2} \frac{dj(x)}{dx} \,.
\ee

The identity (\ref{eq2.21}) proves the assertion just made about the identical
information content in $E(x)$ and $j(x)$.  It also proves several interesting
facts.  First, the location of the minimum of (the exact) $j(x)$ coincides with
that of the minimum of (the exact) $E(x)$ (both of them equivalently defining
the LSO).  Second, the existence of a branch cut singularity
$\propto(x_0-x)^{-1/2}$ in either $j(x)$ or $E(x)$ necessarily implies the
presence of a similar singularity $\propto(x_0-x)^{-1/2}$ (at the same location
$x_0$) in the other function ($E(x)$ or $j(x)$, respectively).  This can be
viewed as a confirmation of our generic assumption of structural stability.
However, this argument also shows the ambiguities present when trying to work
with the energy function.  Indeed, if we assume that, near $x_0$, $j(x)$ can be
expanded as $\varphi(x)(x_0-x)^{-1/2}$, where $\varphi(x)$ is a smooth function,
one finds from Eq.\ (\ref{eq2.21}) that $E(x)=\psi(x)(x_0-x)^{-1/2}+c$, with
some unknown constant $c$.  The lack of knowledge of the constant $c$ (which a
priori depends on the parameter $\nu$) implies that we do not know which
function $(E-c\,(\nu))$ we would square to get a meromorphic function with a
simple pole.  The same reasoning shows another defect of the proposal to
consider the (new) function $e(x)$.  Indeed, when $E(x)\rightarrow\infty$ the
leading term in $e(x)$, Eq.\ (\ref{eq2.12}), will be
$e(x)\sim\frac{1}{4}\nu^2E^4(x)$, which will have a double pole
$\propto\nu^2(x_0-x)^{-2}$, if $j^2(x)$ has a simple pole $\propto(x_0-x)^{-1}$.
For all these reasons, we consider that the ``$j$-method'', Eq.\ (\ref{eq2.19}),
appears as the best way of locating the LSO, within the class of methods dealt
with in this section.

To conclude this section, let us, however, mention another invariant function 
one might wish to consider. This function is the fourth power of the 
dimensionless periastron parameter $1+k=\Phi/(2\pi)$ considered as a function of 
the reduced angular momentum. Indeed, in the test-mass limit, and for circular 
orbits, one knows that \cite{DS88,JS91,DJS1}
\be
\label{eq2.22}
(1+k)^4 = \left(1-\frac{12}{j^2}\right)^{-1} \,.
\ee
We recall that $j^2=12$ is the location of the LSO. Therefore, if we define
$K\equiv(1+k)^4=\biglb(\Phi/(2\pi)\bigrb)^4$ and $y\equiv1/j^2$, we might 
consider
\be
\label{eq2.23}
K_{P_n}(y;\nu) \equiv P_{\ell}^k\,[T_n[K(y;\nu)]] \,,
\ee
with $k+\ell=n$, and our canonical choice of $\ell>0$.  Then, we can take the
pole of $K_{P_n}(y;\nu)$ as estimate of the value of $1/j^2$ at the LSO when
$\nu\ne0$.  We consider, however, that the $j$-method (or any energy method for
that matter) has a better chance of accurately locating the LSO because it
incorporates not only the information that something special (a minimum) takes
place at the LSO, but also the information that the location of this minimum is
a corollary of the presence of a blow up ($j^2\rightarrow\infty$,
$E\rightarrow\infty$) at a point, $x_{\rm pole}$, further away on the real axis.
Therefore, even if the location of $x_{\rm pole}$ is not known very accurately,
one can hope that $x_{\rm LSO}$ will be more robustly determined.

\section{Effective one-body method}

In this section, we shall turn to a rather different method, though one 
which also makes use of the three ideas (i)--(iii) listed at the 
beginning of the previous section. This new method incorporates a fourth 
idea, which has been recently put forward by Buonanno and Damour 
\cite{BD99}. This fourth idea consists in mapping (through the use of 
invariant functions) the real two-body problem we are interested in (two 
masses $m_1$, $m_2$ orbiting around each other) onto an ``effective 
one-body problem'' (one mass $m_0$ moving in some background metric, 
$g_{\alpha \beta}^{\rm effective} (x^{\gamma})$). At the 2PN level, 
Ref.\ \cite{BD99} has shown the possibility of mapping the real two-body 
problem onto geodesic motion in some spherically symmetric metric 
$g_{\alpha\beta}^{\rm effective}(x^{\gamma};\nu)$. It was found 
that, when $\nu\ne0$, $g_{\alpha\beta}^{\rm effective}(x^{\gamma};\nu)$ is a 
smooth deformation of the Schwarzschild metric. The ``mapping 
rules'' between the two problems were motivated by quantum 
considerations:

(i) the adiabatic invariants $I_i = \oint p_i \, dq_i$ (which are 
quantized in units of $\hbar$) were identified in the two problems, and

(ii) the energies are mapped through a function ${\cal E}_{\rm 
effective} = f \, [{\cal E}_{\rm real}]$ which is, a priori, arbitrary, 
and which is determined in the process of matching the two problems.

In other words, the idea is to determine a metric 
$g_{\alpha\beta}^{\text{effective}}$ such that the ``energy levels''
${\cal E}_{\text{effective}}[I_i]$ (i.e.\ the Hamiltonian expressed in action 
variables, or ``Delaunay Hamiltonian'') of the bound states of a particle moving 
in $g_{\alpha\beta}^{\text{effective}}$, are in one-to-one correspondence with 
the two-body bound states:
\be
\label{eq3.1}
{\cal E}_{\rm effective} \, [I_i] = f \, [{\cal E}_{\rm real} \, [I_i]] \,.
\ee
The unknowns of the problem are the numerical coefficients entering a 
spherically symmetric metric
\be
\label{eq3.2}
ds_{\rm eff}^2 = -A(R_{\rm eff})\,dt_{\rm eff}^2
+ \frac{D(R_{\rm eff})}{A(R_{\rm eff})}\,dR_{\rm eff}^2
+ R_{\rm eff}^2\,
(d\theta_{\rm eff}^2+\sin^2\theta_{\rm eff}\,d\varphi_{\rm eff}^2) \,,
\ee
namely,
\begin{mathletters}
\bea
\label{eq3.3}
A(R) &=& 1 + a_1\,\frac{GM_0}{R} + a_2\left(\frac{GM_0}{R}\right)^2
+ a_3\left(\frac{GM_0}{R}\right)^3 + a_4\left(\frac{GM_0}{R}\right)^4
+ \cdots \,,
\\[2ex]
\label{eq3.4}
D(R) &=& 1 + d_1\,\frac{GM_0}{R} + d_2\left(\frac{GM_0}{R}\right)^2
+ d_3\left(\frac{GM_0}{R}\right)^3 + \cdots, 
\eea
\end{mathletters}
and the coefficients entering the energy-map $f$ (here written for the 
``non-relativistic'' energies ${\cal E}_{\rm eff}^{\rm NR} \equiv {\cal 
E}_{\rm eff} - m_0$, ${\cal E}_{\rm real}^{\rm NR} \equiv {\cal E}_{\rm 
real} - M$):
\be
\label{eq3.5}
\frac{{\cal E}_{\rm eff}^{\rm NR}}{m_0} = \frac{{\cal E}_{\rm real}^{\rm 
NR}}{\mu} \left[ 1 + \alpha_1 \, \frac{{\cal E}_{\rm real}^{\rm 
NR}}{\mu} + \alpha_2 \left( \frac{{\cal E}_{\rm real}^{\rm NR}}{\mu} 
\right)^2 + \alpha_3 \left( \frac{{\cal E}_{\rm real}^{\rm NR}}{\mu} 
\right)^3 + \cdots \right] \,.
\ee

As discussed in Ref.\ \cite{BD99} it is natural to require that the effective 
mass $m_0$ be exactly equal to the usual non-relativistic effective mass 
$\mu\equiv m_1\,m_2/(m_1+m_2)$. One can also (by convention) choose the mass 
$M_0$ entering the effective metric to be $M_0 \equiv M = m_1 + m_2$. With these 
choices the Newtonian limit tells us that the 
first coefficient in $A(R)=-g_{00}$ is simply $a_1=-2$. The 1PN 
level then involves the coefficients $a_2$, $d_1$, and $\alpha_1$, while 
the 2PN and 3PN levels involve $(a_3,d_2,\alpha_2)$ and $(a_4,d_3,\alpha_3)$, 
respectively. In other words, at each PN level, we only 
have three arbitrary coefficients to play with. This seems to be quite a 
small number of degrees of freedom, compared to the many possible 
coefficients which can enter the PN-expansion of the Delaunay 
Hamiltonian. [The Delaunay Hamiltonian was determined at the 2PN level 
in Ref.\ \cite{DS88}, and at the 3PN level in Ref.\ \cite{DJS1}.] In order to 
clarify the number of independent equations to be satisfied when mapping 
the real problem onto the effective one, let us consider a 
generic\footnote{We use the information that the leading kinetic terms 
in a PN Hamiltonian are given by the expansion of the free Hamiltonian 
$\sqrt{{\mathbf p}_1^2 + m_1^2} + \sqrt{{\mathbf p}_2^2 + m_2^2}$, so 
that, at the $n$PN level, they are $\propto {\mathbf p}^{2(1+n)}$ 
without dependence on ${\mathbf n}\equiv{\mathbf q}/q$.} PN-expanded 
Hamiltonian, with the symbolic structure
\be
\label{eq3.6}
\widehat{H}_{n{\rm PN}}^{\rm NR}({\mathbf q},{\mathbf p}) = p^{2(n+1)}
+ \frac{1}{q} \left[p^{2n}+p^{2n-2}(np)^2+\cdots+(np)^{2n}\right]
+ \frac{1}{q^2} \left[p^{2(n-1)}+\cdots+(np)^{2(n-1)}\right]
+ \cdots + \frac{1}{q^{n+1}} \, .
\ee
Here, we consider the reduced Hamiltonian
$\widehat{H}^{\rm NR}\equiv H^{\rm NR}/\mu$, in the center-of-mass frame, as a 
function of the reduced canonical variables
${\mathbf p}\equiv{\mathbf p}_1/\mu=-{\mathbf p}_2/\mu$,
${\mathbf q}\equiv({\mathbf x}_1-{\mathbf x}_2)/(GM)$ ($(np)$ denotes
${\mathbf n}\cdot{\mathbf p}$ with ${\mathbf n}\equiv{\mathbf q}/q$). See Ref.\ 
\cite{DJS1} for details. Note that we follow here \cite{BD99} in denoting by 
${\mathbf q}$, ${\mathbf p}$ the original (ADM-like) coordinates. We have 
suppressed all coefficients in Eq.\ (\ref{eq3.6}) to display the structure. What 
is important for our present purpose is the total number of coefficients in the 
$n$PN-level Hamiltonian (\ref{eq3.6}). This is easily checked to be:
\be
\label{eq3.7}
C_H (n) = \frac{(n+1)(n+2)}{2} + 1 \,.
\ee
As explained in Ref.\ \cite{BD99}, one way (and the only explicit one) to 
map the real Hamiltonian (\ref{eq3.6}) onto an effective Hamiltonian, 
while keeping the action variables invariant, is to apply a canonical 
transformation, with generating function $\widetilde G (q,p') = q^i \, 
p'_i + G (q,p')$. The most generic generating function that we need to 
consider has the symbolic structure (at the $n$PN level)
\be
\label{eq3.8}
G_{n{\rm PN}}({\mathbf q},{\mathbf p})
= ({\mathbf q}\cdot{\mathbf p}) \left\{ p^{2n} + \frac{1}{q} \left[ p^{2(n-1)}
+ \cdots + (np)^{2(n-1)} \right] + \cdots + \frac{1}{q^n} \right\} \, . 
\ee
Correspondingly to the pure ${\mathbf p}$-dependence of the leading 
kinetic term in (\ref{eq3.6}) we have written here the leading term in 
$G_{n{\rm PN}}$ as $\propto (qp) \, p^{2n}$. [It is easily shown that 
any term of the form $(qp)\,p^{2(n-k)}\,(np)^{2k}$ must have a 
vanishing coefficient.] The number of arbitrary coefficients in the 
$n$PN-level generating function (\ref{eq3.8}) is easily seen to be
\be
C_G (n) = C_H (n-1) = \frac{n (n+1)}{2} + 1 \, . \label{eq3.9}
\ee
Finally, the difference $\Delta (n)$ between the number of equations to 
satisfy, and the number of unknowns (including the 3 basic parameters 
($a_{n+1}$, $d_n$, $\alpha_n$) appearing in the effective metric and the 
energy-map) reads, at the $n$PN level
\be
\Delta (n) = C_H (n) - C_G (n) - 3 = n-2 \, . \label{eq3.10}
\ee
In particular: $\Delta(1)=-1$, which means that requiring a 1PN matching leaves 
one degree of freedom unrestricted. [This freedom was used in \cite{BD99} to 
impose the natural condition $d_1=0$, i.e.\ that the linearized effective metric 
coincides with Schwarzschild.] Then $\Delta(2)=0$, which means that there will 
(barring any degeneracy) be a unique solution at the 2PN level. [This was indeed 
the result of \cite{BD99}.] But $\Delta(3)=+1$, which means that, at the 3PN 
level, there is one more equation to satisfy than the number of free parameters. 
Then the situation would get worse and worse at higher PN levels.

By explicitly performing the matching between the canonically-transformed
Hamiltonian and (modulo the energy map (\ref{eq3.5})) the effective Hamiltonian
of a point particle moving in some $g_{\mu\nu}^{\rm eff}$, we have established
(details will be given below) that, if we follow Ref.\ \cite{BD99} in imposing
the natural condition $d_1=0$ at the 1PN level, there are, indeed, $C_H (3)=11$
linearly {\it independent} equations, for $C_G(3)+3=10$ unknowns, to be
satisfied at the 3PN level.  [No miracle occurred!]

At face value, this looks like bad news for the idea of the ``effective 
one-body approach''. However, we think that there are acceptable ways to 
rescue this approach. A first cure would be to take advantage of the 
fact that the total number of equations at the three first PN levels (1, 
2, and 3) is exactly equal to the number of unknowns (in other words 
$\Delta(1)+\Delta(2)+\Delta(3)=-1+0+1=0$). Therefore if 
we relax the (not really necessary) constraint $d_1 = 0$, there will be 
a unique 3PN-accurate effective metric $g_{\mu \nu}^{\rm eff}$ (and a unique 
energy mapping (\ref{eq3.5})) satisfying the necessary constraints. We 
have verified that this is true and, for completeness, we give this 
unique solution in Appendix A. But, we do not want to take this solution 
too seriously for the following reasons: (i) it does not look natural to 
have to wait to know the 3PN Hamiltonian to determine the 1PN and 2PN 
effective metrics; (ii) this solution looks more complicated than the 
3PN Hamiltonian itself; and (iii) this trick is not expected to be 
sufficient to ensure the existence of solutions at higher PN levels. 
(Indeed, $\Delta(n)=n-2$ continues to increase.)

We propose therefore to consider a second (more radical, and simpler) 
way to cure the problem. Indeed, we have to face the fact that there is 
(probably) nothing deep in the effective-one-body approach. After all, 
it is just a somewhat artificial way of mapping the complicated two-body 
dynamics on a simpler one-body dynamics. There is no reason to assume 
that the one-body dynamics can, to all orders, be considered as 
equivalent to a simple geodesic motion. Let us recall that, in quantum 
mechanics\footnote{We recall that the origin of the effective-one-body approach 
lies in the quantum electrodynamics of two-charge systems, see \cite{BIZ70}.}, 
geodesic motion means a simple Klein-Gordon Lagrangian
\be
\label{eq3.11}
{\cal L}_{\rm eff} = - \sqrt{g_{\rm eff}} \left( (\nabla \varphi)^2 + 
m_0^2 \, \varphi^2 \right) \quad \hbox{with} \quad (\nabla \varphi)^2 = 
g_{\rm eff}^{\alpha \beta} \, \partial_{\alpha} \, \varphi \, 
\partial_{\beta} \, \varphi \,.
\ee
It is well-known that effective actions generally include, at higher orders, 
some higher-derivative terms: for instance of the type $(\Box_g\,\varphi)^2$, 
$(\nabla\varphi)^4$, etc. Coming back to the classical limit, i.e.\ to the 
Hamilton-Jacobi equation (obtained by considering that
$\varphi(x)=\exp\,(i\,S(x)/\hbar)$ with $\hbar\rightarrow0$), we should 
correspondingly expect that, at higher orders, the effective one-body 
``Hamilton-Jacobi'' equation be of the generalized form
\be
\label{eq3.12}
0 = m_0^2 + g_{\rm eff}^{\alpha \beta} (x) \, p_{\alpha} \, p_{\beta} + 
A^{\alpha \beta \gamma \delta} (x) \, p_{\alpha} \, p_{\beta} \, 
p_{\gamma} \, p_{\delta} + \cdots
\ee
with $p_{\alpha}=\partial\,S(x)/\partial\,x^{\alpha}$. If we were to use a 
Lagrangian formulation, the general form (\ref{eq3.12}) would correspond to an 
action $S=-m_0\int ds_{\rm eff}\left[1+ A_{\alpha\beta\gamma\delta}\,(x)\, 
u^{\alpha}\,u^{\beta}\,u^{\gamma}\,u^{\delta}+\cdots\right]$ with
$u^{\alpha}=dx^{\alpha}/ds_{\rm eff}$, i.e.\ to a general (perturbative) Finsler 
structure.

If we use perturbatively the lowest-order ``on shell'' condition (i.e.\
$m_0^2+g_{\rm eff}^{\alpha\beta}\,p_{\alpha}\,p_{\beta}\simeq0$) we can (for 
instance) eliminate the presence of the energy ${\cal E}_{\rm eff}=-p_0$ in the 
quartic (and higher) terms in (\ref{eq3.12}). In other words, we can restrict 
ourselves to considering purely spatial higher-order tensors 
$A^{\alpha\beta\gamma\delta}(x)=A^{ijk\ell}(x)$, etc. Dimensional analysis shows 
that the quartic terms $({\cal O}\,({\mathbf p}^4))$ in Eq.\ (\ref{eq3.12}) 
which can enter at the 3PN level must have a $q$-dependence of the type 
$A^{ijk\ell}(x)\sim q^{-2}$. Finally, we are led to consider, at the 3PN 
accuracy, (after solving Eq.\ (\ref{eq3.12}) with respect to the effective 
energy ${\cal E}_{\rm eff}=-p_0$) a generalized effective Hamiltonian of the 
form
\be
\label{eq3.13}
\widehat{H}_{\rm eff}^{\rm R}({\mathbf q}',{\mathbf p}')
= \sqrt{A (q') \left[ 1 + {\mathbf p}'^2 + \left( \frac{A(q')}{D(q')} - 
1 \right) ({\mathbf n}' \cdot {\mathbf p}')^2 + \frac{1}{q'^2} \biglb( 
z_1 ({\mathbf p}'^2)^2 + z_2 \, {\mathbf p}'^2 ({\mathbf n}' \cdot {\mathbf 
p}')^2 + z_3 ({\mathbf n}' \cdot {\mathbf p}')^4 \bigrb) \right]} \,, 
\ee
where the quartic (in ${\mathbf p}'$) terms come from the 
$A^{\alpha\beta\gamma\delta}$ coupling and modify the normal ``geodesic'' 
Hamiltonian $\sqrt{-g_{00}^{\rm eff}(1+g_{\rm eff}^{ij}\,p'_i\,p'_j)}$. In 
anticipation of the need to transform (via a canonical transformation) the 
original coordinates $({\mathbf q},{\mathbf p})$ of the real (reduced) 
Hamiltonian into the coordinates of the effective dynamics, we have denoted the 
latter by $({\mathbf q}',{\mathbf p}')$.

This procedure introduces three new arbitrary degrees of freedom at the 3PN
level:  $z_1$, $z_2$, and $z_3$.  It is clear that it now becomes possible to
map the real dynamics on the generalized effective dynamics (\ref{eq3.13}).
This becomes, in fact, possible in many ways.  As we are primarily interested in
(quasi-)circular orbits we shall find convenient to consider only the simple
case where $z_1=z_2=0$, i.e.\ to use only $z_3\ne0$ as new degree of freedom.
[This degree of freedom then disappears in the discussion of circular orbits,
which can then be considered as following essentially from a ``geodesic''
dynamics.]

An important feature of our proposal (\ref{eq3.12}) is that it is 
clearly general enough to allow for the existence of solutions at 
arbitrary PN orders. For instance, at 4PN we would have the freedom to 
introduce either arbitrary sextic terms, $q'^{-2} \, [p'^6 + p'^4 \, 
(n' \, p')^2 + \cdots ]$, or a modification of the quartic terms: $q'^{-3} 
\, [p'^4 + \cdots ]$. This is more freedom than is needed to compensate 
for $\Delta (4) = 4-2 = +2$, Eq.\ (\ref{eq3.10}). We can also clearly 
arrange things so that circular orbits always follow from a ``geodesic'' 
dynamics. The only somewhat unsatisfactory feature of the proposal 
(\ref{eq3.12}) is that we cannot offer any principle for determining a 
priori the structure of the ``post geodesic'' terms. In particular, it 
would have been nice to say that (as is often the case in effective 
actions) the additional terms are somehow generated by the leading term, 
i.e.\ by the effective metric. We have in mind here relations of the type:
$$
A_{\alpha \beta \gamma \delta} = \lambda_1 \, R_{\alpha \beta} \, 
R_{\gamma \delta} + \lambda_2 \, \nabla_{\alpha \beta} \, R_{\gamma 
\delta} + \lambda_3 \, \nabla_{\alpha \beta \gamma \delta} \, R + \cdots,
$$
where $R_{\alpha \beta}$ is the Ricci tensor of $g_{\alpha \beta}^{\rm 
eff}$. However, we have checked that such ``geometrical'' tensors do not 
yield possible 3PN corrections, but only much smaller contributions 
(starting at 5PN).

We shall give in the next section the details of the computation of the 
coefficients entering $A(q')$, $D(q')$, the generating function 
$G_{3{\rm PN}}$, the energy-mapping function and $z_3$. For the time 
being, let us only stress some conceptual points. First, it is 
remarkable that the {\it unique} solution of the 3PN real 
$\rightleftarrows$ effective matching problem leads to the simple value
\be
\label{eq3.14}
\alpha_3 = 0 \,,
\ee
for the 3PN parameter entering the energy-mapping (\ref{eq3.5}). Ref.\ 
\cite{BD99} had already found that $\alpha_2=0$ at 2PN, and $\alpha_1=\nu/2$ at 
1PN. These values correspond exactly to
\be
\frac{{\cal E}_{\rm eff}^R}{m_0} = \varphi \, (s_{\rm real}) = 
\frac{({\cal E}_{\rm real}^R)^2 - m_1^2 - m_2^2}{2 m_1 m_2} \, . 
\label{eq3.15}
\ee
We find remarkable that the simplest, symmetric function of the Mandelstam 
invariant $s_{\rm real}=({\cal E}_{\rm real}^R)^2$ which reduces to
${\cal E}_0/m_0$ in the test-mass limit turns out to define the unique 
energy-map needed to link the real 3PN dynamics to the effective dynamics. We 
interpret this as a good sign for our generalized dynamics (\ref{eq3.12}). [By 
contrast, the other proposal of relaxing the natural constraint $d_1=0$ leads to 
a very complicated energy-map with $\alpha_1\ne\nu/2$, $\alpha_2\ne0$, and 
$\alpha_3\ne0$, see Appendix A.]

Let us now consider the consequences of the effective-one-body approach 
for the determination of the LSO. For the case of circular orbits the 
effective-one-body approach boils down to saying that the real energy is 
the following function of the effective-one,
\be
\label{eq3.16}
{\cal E}_{\rm real}^R = M \sqrt{1 + 2\nu\frac{{\cal E}_{\rm eff}^R - \mu}{\mu}}
\ee
(as obtained by inverting Eq.\ (\ref{eq3.15})) and that the effective 
energy along circular orbits is the square-root of the minimum value of 
a certain ``effective radial potential'':
\be
\frac{{\cal E}_{\rm eff}^R}{\mu} = \sqrt{[W_j(q')]_{\rm min}} \,, 
\label{eq3.17}
\ee
where $W_j(q')$ is obtained from (the square of) Eq.\ (\ref{eq3.13})
by setting ${\mathbf n}'\cdot{\mathbf p}'=0$ (and ${\mathbf p}'^2 =
({\mathbf n}'\times{\mathbf p}')^2+({\mathbf n}'\cdot{\mathbf p}')^2=j^2/q'^2$)
\be
\label{eq3.18}
W_j(q') = A(q')
\left( 1 + \frac{j^2}{q'^2} + z_1 \, \frac{j^4}{q'^6} \right) \,.
\ee
As said above, we shall assume (as is always possible) that $z_1=0$, 
so that the effective potential has the usual ``Schwarzschild-like'' 
value $-g_{00}(q')\,(1+j^2/q'^2)$. The value of the metric 
coefficient $A(q')=-g_{00}(q')$ is obtained, at the 3PN level, as 
a perturbative expansion in $1/q'=GM/R_{\rm eff}$:
\be
\label{eq3.19}
A(q') = 1 - \frac{2}{q'} + \frac{2\nu}{q'^3} + \frac{a_4(\nu)}{q'^4}
+ {\cal O}\left(\frac{1}{q'^5}\right) \,. 
\ee
Note that, finally, in this approach the entire effect of the 3PN dynamics (for 
circular orbits) is contained in the sole coefficient $a_4(\nu)$ (whose value 
will be discussed in the next section).

In Ref.\ \cite{BD99} the metric coefficient $A(q')$ was used (at the 2PN level) 
as a simple truncated Taylor expansion: $A_{2{\rm PN}}(q')=1-2/q'+2\nu/q'^3$. 
This simple-minded approach is not adequate for dealing with the 3PN level. 
Indeed, we shall see in next section that the coefficient $a_4(\nu)$ is positive 
and can be relatively large. In keeping with the spirit of the present work 
where we systematically try to use adequate resummation methods to improve the 
convergence of PN expansions, we shall {\it define} our effective-one-body 
radial potential, at the $n$PN accuracy (expressed in terms of the convenient 
variable $u\equiv{1/q'}$)
\be
\label{eq3.20}
W_j^{P_n}(u) = A_{P_n}(u) \left(1+j^2\,u^2\right)
\ee
by using a suitable Pad\'e approximant of Eq.\ (\ref{eq3.19}):
\be
\label{eq3.21}
A_{P_n}(u) \equiv P_{\ell}^k \left[ T_{n+1} [A(u)] \right] \, , 
\ee
with $k+\ell=n+1$ (because it is $q'^{-n-1}$ which corresponds to the $n$PN 
level) and, now the constraint $k>0$ (instead of $\ell>0$ as above), because we 
want to factor a zero of $A(u)$ (and no longer a pole). The Pad\'e improvement 
of $A(u)$ is really needed (and makes a difference) only at the 3PN level. We 
have found that the most robust Pad\'es (under variation of the Taylor 
coefficients) are defined by taking $k=1$ and $\ell=n$.

Summarizing the present method: The effective radius $q'$ of circular orbits is 
obtained as a function of the reduced angular momentum $j$ by looking for the 
minimum of the radial potential (\ref{eq3.20}), where $u\equiv{1/q'}$ and where 
the Pad\'e-improved function $A$ is given by Eq.\ (\ref{eq3.21}), with $k=1$ and 
$\ell=n$. For each value of $j$ above some threshold $j_{\rm min}$, $W_j(u)$ 
admits a (unique) minimum $u_*(j)$. {From} this one then determines the 
effective-energy (\ref{eq3.17}), and then the real one (\ref{eq3.16}), namely
\be
\label{eq3.22}
{\cal E}_{\rm real}^{R}(j)
= M \sqrt{ 1+2\nu \left[\sqrt{W_j(u_*(j))}-1\right] } \,.
\ee
The real circular orbital frequency corresponding to $j$ is then given by 
using the identity (\ref{eq2.20}). This yields
\be
\label{eq3.23}
GM\,\omega_{\rm real}(j) = \frac{j\,u_*^2(j)\ \sqrt{A_{P_n}(u_*(j))}} 
{\sqrt{1+j^2\,u_*^2(j)}\ \sqrt{1+2\nu\left[\sqrt{W_j(u_*(j))}-1\right]}} \,.
\ee
Note that in this approach $j\equiv j_{\rm real}\equiv j_{\rm effective}$. 
Finally, the LSO is invariantly defined as the minimum value of $j$,
$j_{\rm LSO}=j_{\rm min}$, for which $W_j(u)$ admits a local minimum. When 
$j<j_{\rm LSO}$, $W_j(u)$ has no local minimum and there are no (stable or 
unstable) circular orbits (see, e.g., Fig.\ 1 of Ref.\ \cite{BD99}). If one is 
only interested in locating the LSO (as a function of $\nu$) it suffices to 
look for the existence of an inflection point of $W_j(u)$, i.e.\ to solve the 
simultaneous equations $\partial\,W_j(u)/\partial\,u=0$ and 
$\partial^2\,W_j(u)/\partial\,u^2=0$.

\section{Results}

Let us now give the details of the application, at the 3PN level, of the methods 
explained above. We follow the order of presentation given in the previous 
two sections. In what follows, we use the  results of Ref.\ \cite{DJS1} 
for the dynamical invariants of the 3PN two-body Hamiltonian. We recall that the 
3PN Hamiltonian derived in Ref.\ \cite{JS98} contained two ambiguous parameters 
$\oms$ and $\omk$ (these ambiguities arise because of the need to regularize 
badly divergent integrals \cite{JS98,JS99,DJS1}), and that all the dynamical 
invariants involve only the combination 
$\sigma\,(\nu)\equiv\oms\,\nu+\omk\,\nu^2$ \cite{DJS1}.

Recently the `kinetic' ambiguity parameter $\omk$ was uniquely determined 
\cite{DJS2} (see also \cite{BF00}) to be $\omk=41/24$, so that
$\sigma\,(\nu)=\oms\,\nu+\frac{41}{24}\,\nu^2$ and the remaining 3PN ambiguity 
is embodied in the product $\oms\,\nu$. As discussed in the Appendix A of 
\cite{DJS1} one expects (both by judging from the other coefficients, and by 
looking at some of the sources of ambiguity) that $\oms$ 
varies in the range
\be
\label{oms}
-10 \leq \oms \leq +10.
\ee

\subsection{$e$-method}

For self-containedness let us quote the results obtained in our previous paper 
\cite{DJS1} for the link between the energy and the $x$-variable, Eq.\ 
(\ref{eq2.1}). The original energy function $E$, Eq.\ (\ref{eq2.2}), admits the 
PN expansion (\ref{eq2.4}) with coefficients
\begin{mathletters}
\label{e01}
\begin{eqnarray}
E_1(\nu) &=& -\frac{1}{12} (9+\nu),
\\[2ex]
E_2(\nu) &=& -\frac{1}{24} (81-57\nu+\nu^2),
\\[2ex]
E_3(\nu) &=& -\frac{10}{3} \biglb(w_1(\nu)-\oms\,\nu\bigrb),
\end{eqnarray}
\end{mathletters}
where
\be
\label{e02}
w_1(\nu) \equiv \frac{405}{128}
+ \frac{1}{64}\bigg(41\pi^2-\frac{6889}{6}\bigg)\nu + \frac{31}{64}\nu^2
+ \frac{7}{3456}\nu^3.
\ee
Correspondingly to this expansion, the new energy function $e$, Eq.\ 
(\ref{eq2.12}), admits the expansion (\ref{eq2.14}) with coefficients
\begin{mathletters}
\label{e03}
\begin{eqnarray}
e_1(\nu) &=& -\bigg(1+\frac{1}{3}\nu\bigg),
\\[2ex]
e_2(\nu) &=& -\bigg(3-\frac{35}{12}\nu\bigg), 
\\[2ex]
e_3(\nu) &=& -\frac{10}{3}\biglb(w_2(\nu)-\oms\,\nu\bigrb),
\end{eqnarray}
\end{mathletters}
where
\be
\label{e04}
w_2(\nu) \equiv \frac{27}{10}
+ \frac{1}{16}\bigg(\frac{41}{4}\pi^2-\frac{4309}{15}\bigg)\nu
+ \frac{103}{120}\nu^2 - \frac{1}{270}\nu^3.
\ee
The 2PN and 3PN Pad\'es of $e(x)$ are given by (see Ref.\ \cite{DIS} for the 2PN 
case)
\begin{mathletters}
\label{e05}
\begin{eqnarray}
e_{P_2}(x) &\equiv& P^1_1\left[T_2[e(x)]\right] = -x
\frac{1+\frac{1}{3}\nu-\big(4-\frac{9}{4}\nu+\frac{1}{9}\nu^2\big)x}
{1+\frac{1}{3}\nu-\big(3-\frac{35}{12}\nu\big)x},
\\[2ex]
e_{P_3}(x) &\equiv& P^2_1\left[T_3[e(x)]\right] = -x
\frac{1 - \biglb(1+\frac{1}{3}\nu+w_3(\nu)\bigrb)x
- \biglb(3-\frac{35}{12}\nu-\big(1+\frac{1}{3}\nu\big)w_3(\nu)\bigrb)x^2}
{1-w_3(\nu)x},
\end{eqnarray}
\end{mathletters}
where
\be
\label{e06}
w_3(\nu) \equiv \frac{40}{36-35\nu}\biglb(w_2(\nu)-\oms\,\nu\bigrb).
\ee
The corresponding $x$-location of the $e$-LSO (minimum of $e(x)$) can be 
written down analytically only at the 2PN level \cite{DIS}:
\be
\label{e07}
6\,x_{\rm LSO}^{e_{P_2}}(\nu)
= \frac{1+\frac{1}{3}\nu}{1-\frac{35}{36}\nu}
\left( 2 - \frac{1+\frac{1}{3}\nu}
{\sqrt{1-\frac{9}{16}\nu+\frac{1}{36}\nu^2}} \right).
\ee
Note that $6\,x_{\rm LSO}^{e_{P_2}}\left(\frac{1}{4}\right)=1.1916$, which 
means that the $e_{P_2}$-predicted value of the orbital frequency at the LSO 
differs from the ``Schwarzschild value'',
$GM\omega_{\rm LSO}^{\rm Schw}=(x_{\rm LSO}^{\rm Schw})^{3/2}$ with
$x_{\rm LSO}^{\rm Schw}=1/6$, by the factor
\be
\label{e08}
\widehat{\omega}_{\rm LSO}
\equiv \frac{\omega_{\rm LSO}}{\omega_{\rm LSO}^{\rm Schw}}
= (6\,x_{\rm LSO})^{3/2},
\ee
which is about $1.3007$ in the present case. We will use 
$\widehat{\omega}_{\rm LSO}$ as our main tracer of the ``observable'' location 
of the LSO. It is important to note from the start that (as emphasized in 
\cite{DIS}) the $e$-method predicts (at 2PN) that the orbital frequency at the 
LSO is {\it larger} than the ``Schwarzschild value''. The corresponding results, 
at 3PN, are exhibited in Table \ref{tab:lso}. Let us only note here that the 
tendency to get $\widehat{\omega}_{\rm LSO}>1$ seems confirmed, at the 3PN 
level, rather independently of the value of the ambiguity parameter $\oms$.

Once the value of $x_{\rm LSO}(\nu)$ is determined (analytically or numerically) 
one can compute the corresponding value of the (real) reduced binding energy 
$E$, Eq.\ (\ref{eq2.2}). It is obtained by solving Eq.\ (\ref{eq2.12}) in terms 
of ${\cal E}^R\equiv M+\mu\,E$. The solution is explicitly given by
\be
\label{e09}
E(x) = \frac{1}{\nu}
\left[ \sqrt{1+2\nu\left(\sqrt{1+e(x)}-1\right)} - 1 \right].
\ee
Then, knowing $E(x)$ we can also compute the value of the reduced angular 
momentum $j$ by integrating the identity (\ref{eq2.21}). Integrating 
Eq.\ (\ref{eq2.21}) by parts yields
\be
\label{e10}
j(x) = -2x^{-1/2} \frac{dE(x)}{dx} + 2 \int_0^x d\bar{x}\, {\bar x}^{-1/2} \, 
\frac{d^2E(\bar x)}{d{\bar x}^2} \,,
\ee
where we have incorporated the information that $j(x)\sim x^{-1/2}$ when 
$x\to0$ (i.e.\ in the limit of very wide circular orbits, described by 
Newtonian dynamics). By applying this result to $x=x_{\rm LSO}(\nu)$, one 
finally gets the value of $j_{\rm LSO}(\nu)$. The results so obtained by the 
$e$-method are shown in Table \ref{tab:lso}.

\begin{table}
\caption{ Equal-mass ($\nu=1/4$) binary-system LSO parameters obtained by means 
of different methods. The reduced binding energy $E_{\rm LSO}$, Eq.\ 
(\ref{eq2.2}), and the reduced angular momentum $j_{\rm LSO}$, Eq.\ 
(\ref{eq2.17}), are divided by their ``Schwarzschild values'':
$\vert{E_{\rm LSO}^{\rm Schw}}\vert=1-\frac{1}{3}\sqrt{8}$ and
$j_{\rm LSO}^{\rm Schw}=\sqrt{12}$; the dimensionless orbital frequency 
$\widehat{\omega}_{\rm LSO}$ is defined in Eq.\ (\ref{e08}). The line denoted by 
`BD' reports the 2PN results obtained in Ref.\ [13]. }
\label{tab:lso}
\begin{center}
\begin{tabular}{c|ccccc|ccccc|ccccc}
& \multicolumn{5}{c|}{$E_{\rm LSO}/\vert{E_{\rm LSO}^{\rm Schw}}\vert$}
& \multicolumn{5}{c|}{$j_{\rm LSO}/j_{\rm LSO}^{\rm Schw}$}
& \multicolumn{5}{c}{$\widehat{\omega}_{\rm LSO}$} \\
& & &\multicolumn{3}{c|}{3PN}
& & &\multicolumn{3}{c|}{3PN}
& & &\multicolumn{3}{c}{3PN} \\
method & 1PN & 2PN &\multicolumn{3}{c|}{$\oms$}
& 1PN & 2PN &\multicolumn{3}{c|}{$\oms$}
& 1PN & 2PN &\multicolumn{3}{c}{$\oms$} \\ \cline{4-6}\cline{9-11}\cline{14-16}
& & &$-10$ & 0 & 10
& & &$-10$ & 0 & 10
& & &$-10$ & 0 & 10
\\ \hline
$e$-method
& $-$
& $-1.142$
& $-1.015$
& $-1.369$
& $-1.636$
   & $-$
   & 0.956
   & 0.994
   & 0.904
   & 0.857
& $-$
& 1.301
& 0.998
& 1.969
& 2.870 \\
$j$-method
& $-0.973$
& $-1.091$
& $-1.039$
& $-1.322$
& $-1.842$
   & 1.014
   & 0.970
   & 0.986
   & 0.913
   & 0.833
& 0.960
& 1.173
& 1.060
& 1.774
& 3.915 \\
eff.\ method
& $-1.007$
& $-1.048$
& $-1.042$
& $-1.168$
& $-1.212$
   & 1.000
   & 0.983
   & 0.985
   & 0.946
   & 0.934
& 1.015
& 1.075
& 1.064
& 1.297
& 1.383 \\
BD, Ref.\ \cite{BD99}
& $-1.007$
& $-1.050$
& $-$
& $-$
& $-$
   & 1.000
   & 0.983
   & $-$
   & $-$
   & $-$
& 1.015
& 1.079
& $-$
& $-$
& $-$ \\
$k$-method
& $-$
& $-$
& $-$
& $-$
& $-$
   & 1.000
   & 0.980
   & 0.974
   & 0.955
   & 0.936
& $-$
& $-$
& $-$
& $-$
& $-$ \\
\end{tabular}
\end{center}
\end{table}

\subsection{$j$-method}

In this approach the basic PN expansion we consider is that of $1/j^2(x)$. It 
reads (cf.\ Eq.\ (5.11) in Ref.\ \cite{DJS1})
\be
\label{j0a}
\frac{1}{j^2(x)} = x \left[ 1 - \frac{1}{3}(9+\nu)x + \frac{25}{4}\nu x^2
- \frac{16}{3}\biglb(w_4(\nu)-\oms\,\nu\bigrb)x^3 \right],
\ee
where
\be
\label{j0b}
w_4(\nu) \equiv \frac{1}{64}\bigg(41\pi^2-\frac{5269}{6}\bigg)\nu
+ \frac{61}{64}\nu^2 - \frac{1}{432}\nu^3.
\ee
{From} Eq.\ (\ref{j0a}) one gets
\be
\label{j01}
j^2(x) = \frac{1}{x} \left[ 1  + \frac{1}{3}(9+\nu)\,x
+ \frac{1}{36}(36-\nu)(9-4\nu)\,x^2
+ \frac{16}{3}\biglb(w_5(\nu)-\oms\,\nu\bigrb)\,x^3 \right],
\ee
where
\be
\label{j02}
w_5(\nu) \equiv \frac{81}{16} + 
\frac{1}{64}\bigg(41\pi^2-\frac{7321}{6}\bigg)\nu
+ \frac{23}{64}\nu^2 + \frac{1}{216}\nu^3.
\ee

\begin{figure}
\centerline{\epsfig{file=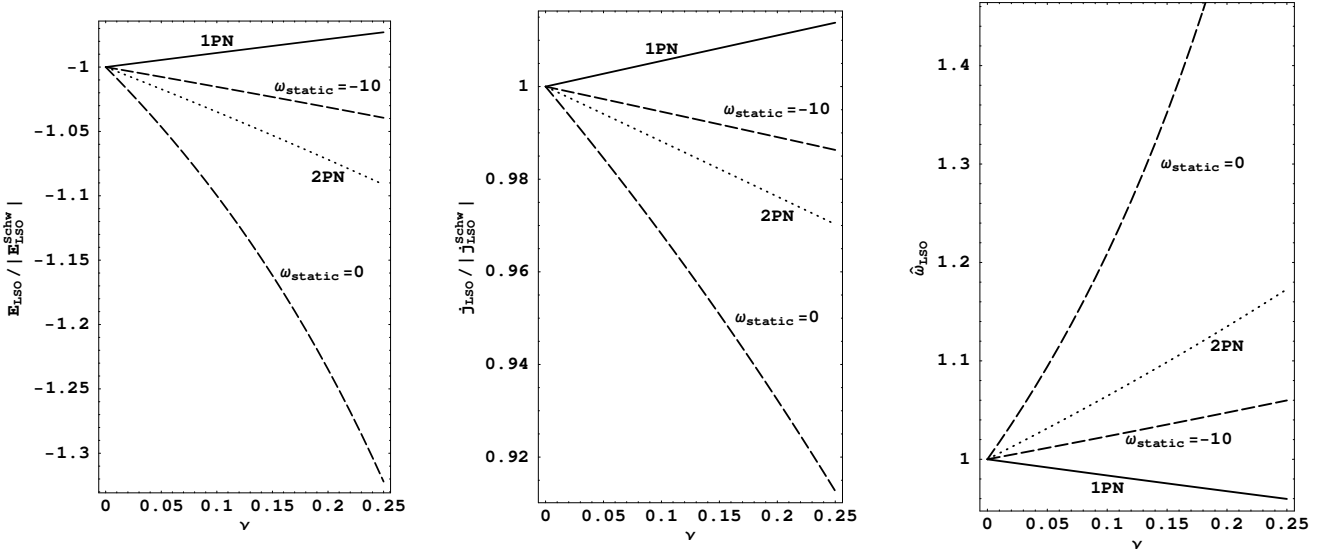,viewport=0 550 612 792}}
\caption{\label{fig:js}
Binary-system LSO parameters obtained by means of the $j$-method as functions 
of the symmetric mass-ratio $\nu$. The reduced binding energy $E_{\rm LSO}$, 
Eq.\ (\ref{eq2.2}), and the reduced angular momentum $j_{\rm LSO}$, Eq.\ 
(\ref{eq2.17}), are divided by their ``Schwarzschild values'':
$\left\vert{E_{\rm LSO}^{\rm Schw}}\right\vert=1-\frac{1}{3}\sqrt{8}$ and
$j_{\rm LSO}^{\rm Schw}=\sqrt{12}$; the dimensionless orbital frequency 
$\widehat{\omega}_{\rm LSO}$ is defined in Eq.\ (\ref{e08}). The lines shown in 
the plots correspond to different PN orders of approximation: 1PN (solid), 2PN 
(dotted), and 3PN (dashed).}
\end{figure}

As explained above we construct the following sequence of near-diagonal Pad\'es 
of $j^2(x)$:
\begin{mathletters}
\label{j03}
\bea
j_{P_1}^2(x) &\equiv& P_1^0\left[T_1[j^2(x)]\right]
= \frac{1}{x\biglb(1-\left(3+\frac{1}{3}\nu\right)x\bigrb)},
\\[2ex]
j_{P_2}^2(x) &\equiv& P_1^1\left[T_2[j^2(x)]\right]
= \frac{1+\frac{1}{9}\nu+\frac{25}{12}\nu x}
{x\biglb(1+\frac{1}{9}\nu-\left(3-\frac{17}{12}\nu+\frac{1}{27}\nu^2\right)x
\bigrb)},
\\[2ex]
j_{P_3}^2(x) &\equiv& P^2_1\left[T_3[j^2(x)]\right]
= \frac{1+\biglb(3+\frac{1}{3}\nu-w_6(\nu)\bigrb) x
+\biglb(9-\frac{17}{4}\nu+\frac{1}{9}\nu^2-\left(3+\frac{1}{3}\nu\right)w_6(\nu)
\bigrb) x^2}
{x \biglb(1-w_6(\nu)x\bigrb)},
\eea
\end{mathletters}
where
\be
\label{j04}
w_6(\nu) \equiv \frac{192}{(36-\nu)(9-4\nu)}\biglb(w_5(\nu)-\oms\,\nu\bigrb).
\ee
The corresponding $x_{\text{LSO}}$ (now defined as the location of the minimum 
of $j(x)$, or, equivalently, $j^2(x)$) can be written down analytically at 1PN 
and 2PN
\begin{mathletters}
\bea
\label{j05}
6\,x_{\rm LSO}^{j_{P_1}}(\nu) &=& \frac{1}{1+\frac{1}{9}\nu},
\\[2ex]
6\,x_{\rm LSO}^{j_{P_2}}(\nu) &=& \frac{8(9+\nu)}{25\nu}
\left[ \frac{2(9+\nu)}{\sqrt{(36-\nu)(9-4\nu)}}-1 \right].
\eea
\end{mathletters}
Note that while $6\,x_{\rm LSO}^{j_{P_1}}(\nu)$ is very slightly smaller than 1, 
$6\,x_{\rm LSO}^{j_{P_2}}(\nu)$ is (like for the $e_{P_2}$ estimate)
{\em larger} than 1. In particular,
$6\,x_{\rm LSO}^{j_{P_2}}\left(\frac{1}{4}\right)=1.1121$, and the corresponding 
dimensionless frequency is
$\widehat{\omega}_{\rm LSO}^{j_{P_2}}\left(\frac{1}{4}\right)=1.1728$. This 
tendency to get ``larger than Schwarzschild'' frequency at the LSO is confirmed 
by the (numerical) 3PN results which are exhibited in Table \ref{tab:lso} and 
Fig.\ \ref{fig:js}.

Within the present $j$-method, once the value of $x_{\rm LSO}(\nu)$ is 
determined one can compute the corresponding value of the (real) reduced binding 
energy $E$ by integrating the identity (\ref{eq2.21}). Indeed, one can write
\be
\label{j06}
E(x) = \int_0^x d\bar{x}\,\bar{x}^{3/2}\frac{dj(\bar{x})}{d\bar{x}} \,,
\ee
where one has incorporated the boundary condition that $E(x)\to0$ when $x\to0$. 
The results so obtained are shown in Table \ref{tab:lso} and Fig.\ \ref{fig:js}.

\subsection{$k$-method}

For completeness, let us (though it is not among our preferred methods) mention 
some results obtained by using the $k$-method, Eq.\ (\ref{eq2.23}). The PN 
expansion of the function $K(y)$, where $K\equiv(1+k)^4$ and $y\equiv1/j^2$, 
reads (cf.\ Eq.\ (5.27) in Ref.\ \cite{DJS1})
\be
\label{k01}
K(y) = 1 + 12y + 24(6-\nu)y^2 + 24\biglb(w_7(\nu)-\oms\,\nu\bigrb)y^3,
\ee
where
\be
\label{k02}
w_7(\nu) \equiv 72 + \bigg(\frac{41}{64}\pi^2-\frac{128}{3}\bigg)\nu
+ \frac{1}{2}\nu^2.
\ee

As explained above, we construct the following sequence of near-diagonal Pad\'es 
of $K(y)$:
\begin{mathletters}
\label{k03}
\bea
K_{P_1}(y) &\equiv& P_1^0\left[T_1[K(y)]\right] = \frac{1}{1-12y},
\\[2ex]
K_{P_2}(y) &\equiv& P_1^1\left[T_2[K(y)]\right] = \frac{1+2\nu y}{1-2(6-\nu)y},
\\[2ex]
K_{P_3}(y) &\equiv& P^2_1\left[T_3[K(y)]\right]
= \frac{1 + \biglb(12-w_8(\nu)\bigrb)y + 12\biglb(2(6-\nu)-w_8(\nu)\bigrb)y^2}
{1-w_8(\nu)y},
\eea
\end{mathletters}
where
\be
\label{k04}
w_8(\nu) \equiv \frac{1}{6-\nu}\biglb(w_7(\nu)-\oms\,\nu\bigrb).
\ee
Then we take the poles of $K_{P_n}$ as estimates of the value of $1/j^2$ at the 
LSO. The results for equal-mass binaries ($\nu=1/4$) are given in Table I.

\subsection{Effective one-body method}

Let us explain in detail how we implemented the effective one-body 
method. Two methods of implementation were presented in Ref.\ \cite{BD99}. We 
could have used the first one by starting from the 3PN Delaunay 
Hamiltonian given in Ref.\ \cite{DJS1}. However, we found finally as 
convenient (given the existence of good algebraic manipulators) to use 
the second method, which has the advantage of being more informative. 
This second method consists of writing explicitly the equations that have 
to be satisfied by the looked for generating function $G (q,p')$ and 
solving them. Indeed, we look for a canonical transformation between the 
original (quasi-ADM) coordinates $(q,p)$ of the real problem (i.e.\ the 
phase-space coordinates in which was obtained the {\it order-reduced} 
Hamiltonian $H(q,p)$ in \cite{DJS1}), and the ``effective'' coordinates 
$(q',p')$ (i.e.\ the coordinates used in Eq.\ (\ref{eq3.13})). The effect 
of the generating function $G(q,p')$ reads
\be
\label{ef01}
q'^i = q^i + \frac{\partial\,G(q,p')}{\partial\,p'_i},\quad
p_i = p'_i + \frac{\partial\,G(q,p')}{\partial\,q^i}.
\ee
Note that, as is well known, the canonical transformation is defined only 
in implicit form: $q'$ and $p$ being given as functions of $q$ and $p'$. But 
there is, in fact, no need to solve for, e.g., $(q',p')$ as functions of 
$(q,p)$. As the basic idea is anyway to identify the numerical value of, 
say, $H_{\rm eff} (q',p')$ with the numerical value of some (to be 
determined) function $f (H_{\rm real} (q,p))$, we can do this 
identification by expressing both sides in terms of any set of common 
variables, say $q$ and $p'$. Finally we write (using Eq.\ (\ref{eq3.5}))
\be
\label{ef02}
\left[ \widehat{H}_{\rm eff}^{\rm R}\biglb(q'(q,p'),p'\bigrb) \right]^2
= \left\{ 1 + \widehat{H}_{\rm real}^{\rm NR}\biglb(q,p(q,p')\bigrb)
\left[ 1 + \alpha_1 \, \widehat{H}_{\rm real}^{\rm NR}
+ \alpha_2 \left( \widehat{H}_{\rm real}^{\rm NR} \right)^2
+ \alpha_3 \left( \widehat{H}_{\rm real}^{\rm NR} \right)^3 \right] \right\}^2,
\ee
in which the L.H.S.\ is given by the R.H.S.\ of Eq.\ (\ref{eq3.13}) (without 
the square root, because we work with the squared equation), while 
$\widehat{H}_{\rm real}^{\rm NR}(q,p)$ on the R.H.S.\ is the 
order-reduced Hamiltonian of \cite{DJS1}, obtained from the higher-order 
3PN Hamiltonian derived in \cite{JS98}. Both sides of Eq.\ (\ref{ef02}) are 
written in terms of $q$ and $p'$ by using Eq.\ (\ref{ef01}). This procedure has 
been used at the 2PN level in \cite{BD99}, so that we know the values of 
$\alpha_1=\frac{1}{2}\nu$, $\alpha_2=0$, the 2PN values of the metric functions 
$A(q')$ and $D(q')$, as well as the 2PN-accurate generating function $G(q,p')$ 
(see e.g.\ Eqs.\ (6.24)--(6.27) in \cite{BD99}). The new unknowns 
entering at the 3PN level are: $\alpha_3$, $a_4$, Eq.\ (\ref{eq3.3}), 
$d_3$, Eq.\ (\ref{eq3.4}), $z_1$, $z_2$, $z_3$, Eq.\ (\ref{eq3.13}), and 
the 7 arbitrary coefficients $c_1$, \ldots, $c_7$ entering the generic form of 
$G_{3{\rm PN}}$:
\be
\label{ef03}
G_{3{\rm PN}}(q,p') = ({\mathbf q}\cdot{\mathbf p}')
\left[ c_1\,{\mathbf p}'^6 + 
\frac{1}{q} \biglb( c_2\,{\mathbf p}'^4
+ c_3\,{\mathbf p}'^2 ({\mathbf n}\cdot{\mathbf p}')^2
+ c_4 ({\mathbf n}\cdot{\mathbf p}')^4 \bigrb)
+ \frac{1}{q^2} \biglb( c_5 \, {\mathbf p}'^2 + c_6 
({\mathbf n} \cdot {\mathbf p}')^2 \bigrb) + \frac{c_7}{q^3} \right] \,. 
\ee
The basic input for writing these equations is the explicit value of the 
3PN-accurate Hamiltonian \cite{JS98,DJS1,DJS2}
\be
\label{ef04}
\widehat{H}_{\text{real}}^{\text{NR}}({\mathbf q},{\mathbf p})
= \widehat{H}_{\rm N}({\mathbf q},{\mathbf p})
+ \widehat{H}_{\rm 1PN}({\mathbf q},{\mathbf p})
+ \widehat{H}_{\rm 2PN}({\mathbf q},{\mathbf p})
+ \widehat{H}_{\rm 3PN}({\mathbf q},{\mathbf p})\,,
\ee
where
\begin{mathletters}
\label{ef05}
\bea
&\label{33}\dst
\widehat{H}_{\rm N}\left({\bf q},{\bf p}\right) = \frac{\pp}{2} - \frac{1}{q},
&\\[2ex]&\dst
\label{34}
\widehat{H}_{\rm 1PN}\left({\bf q},{\bf p}\right) = \frac{1}{8}(3\nu-1)\ppp^2
- \frac{1}{2}\left[(3+\nu)\pp+\nu\np^2\right]\frac{1}{q} + \frac{1}{2q^2},
&\\[2ex]&\dst
\label{35}
\widehat{H}_{\rm 2PN}\left({\bf q},{\bf p}\right)
= \frac{1}{16}\left(1-5\nu+5\nu^2\right)\ppp^3
+ \frac{1}{8} \left[
\left(5-20\nu-3\nu^2\right)\ppp^2-2\nu^2\np^2\pp-3\nu^2\np^4 \right]\frac{1}{q}
\nonumber&\\[2ex]&\dst
+ \frac{1}{2} \left[(5+8\nu)\pp+3\nu\np^2\right]\frac{1}{q^2}
- \frac{1}{4}(1+3\nu)\frac{1}{q^3},
&\\[2ex]&\dst
\label{36}
\widehat{H}_{\rm 3PN}\left({\bf q},{\bf p}\right)
= \frac{1}{128}\left(-5+35\nu-70\nu^2+35\nu^3\right)\ppp^4
\nonumber&\\[2ex]&\dst
+ \frac{1}{16}\left[
\left(-7+42\nu-53\nu^2-5\nu^3\right)\ppp^3
+ (2-3\nu)\nu^2\np^2\ppp^2
+ 3(1-\nu)\nu^2\np^4\pp - 5\nu^3\np^6
\right]\frac{1}{q}
\nonumber&\\[2ex]&\dst
+\left[ \frac{1}{16}\left(-27+136\nu+109\nu^2\right)\ppp^2
+ \frac{1}{16}(17+30\nu)\nu\np^2\pp + \frac{1}{12}(5+43\nu)\nu\np^4
\right]\frac{1}{q^2}
&\nonumber\\[2ex]&\dst
+\left\{ \left[ -\frac{25}{8} + \left(\frac{1}{64}\pi^2-\frac{335}{48}\right)\nu 
- \frac{23}{8}\nu^2 \right]\pp
+ \left(-\frac{85}{16}-\frac{3}{64}\pi^2-\frac{7}{4}\nu\right)\nu\np^2 
\right\}\frac{1}{q^3}
&\nonumber\\[2ex]&\dst
+ \left[ \frac{1}{8} + \left(\frac{109}{12}-\frac{21}{32}\pi^2+\oms\right)\nu 
\right]\frac{1}{q^4}.
\eea
\end{mathletters}
As explained in Refs.\ \cite{JS99,DJS1,DJS2} and at the beginning of the present 
section the 3PN Hamiltonian contains one dimensionless ambiguity parameter 
$\oms$.

When written explicitly, the constraint equation (\ref{ef02}) (truncated at 3PN 
accuracy) yields a system of 11 equations for the $10+3$ unknowns
$(\alpha_3,a_4,d_3,c_1,\ldots,c_7;z_1,z_2,z_3)$ ($\oms$ being assumed to be 
known). This system can be decomposed into three subsystems. The first subsystem
consists of 5 equations:
\begin{mathletters}
\label{ef08a}
\bea
\alpha_3 + 16c_1 &=& -\nu-3\nu^2-5\nu^3,
\\[2ex]
\alpha_3+2c_1-2c_2 &=& -\frac{1}{8}\nu^2-2\nu^3,
\\[2ex]
6c_1+c_2-3c_3 &=& \frac{17}{16}\nu+\frac{19}{4}\nu^2+\frac{27}{16}\nu^3,
\\[2ex]
3c_3-5c_4 &=& -\frac{3}{2}\nu-\frac{27}{2}\nu^2-\frac{81}{16}\nu^3,
\\[2ex]
c_4 &=& \frac{3}{2}\nu^2+\frac{7}{16}\nu^3.
\eea
\end{mathletters}
The second subsystem contains 3 equations:
\begin{mathletters}
\label{ef08b}
\bea
-3\alpha_3+2c_2-2c_5+z_1 &=& \frac{3}{2}\nu-\frac{9}{4}\nu^2+\frac{19}{8}\nu^3,
\\[2ex]
8c_2+6c_3+4c_5-6c_6+z_2 &=& 
\frac{61}{8}\nu+\frac{11}{2}\nu^2-\frac{11}{2}\nu^3,
\\[2ex]
4c_3+10c_4+8c_6+z_3 &=& -\frac{79}{6}\nu-\frac{55}{3}\nu^2+\frac{39}{8}\nu^3,
\eea
\end{mathletters}
and the third one consists also of 3 equations:
\begin{mathletters}
\label{ef08c}
\bea
2\alpha_3+c_5-c_7 &=& \bigg(\frac{271}{48}+\frac{1}{64}\pi^2\bigg)\nu
+\frac{5}{8}\nu^2-\frac{5}{8}\nu^3,
\\[2ex]
-d_3+4c_5+6c_6+6c_7 &=& \bigg(-\frac{35}{8}-\frac{3}{32}\pi^2\bigg)\nu
-\frac{57}{4}\nu^2+2\nu^3,
\\[2ex]
-2\alpha_3+a_4+2c_7 &=& \bigg(\frac{221}{12}-\frac{21}{16}\pi^2+2\oms\bigg)\nu
+\frac{3}{4}\nu^2+\frac{1}{4}\nu^3.
\eea
\end{mathletters}

The first subsystem, Eqs.\ (\ref{ef08a}), yields 5 linear equations for the 5 
unknowns $c_1$, $c_2$, $c_3$, $c_4$, and $\alpha_3$. It is easily found to have 
a {\em unique} solution, namely
\begin{mathletters}
\label{ef09a}
\bea
\alpha_3 &=& 0,
\\[2ex]
c_1 &=& -\frac{1}{16}(1+3\nu+5\nu^2)\,\nu,
\\[2ex]
c_2 &=& -\frac{1}{16}(1+2\nu-11\nu^2)\,\nu,
\\[2ex]
c_3 &=& -\frac{1}{24}(12+48\nu+23\nu^2)\,\nu,
\\[2ex]
c_4 &=& \frac{1}{16}(24+7\nu)\,\nu^2.
\eea
\end{mathletters}
As already mentioned above, note the remarkably simple result $\alpha_3=0$ 
(which confirms that the energy map takes the nice form (\ref{eq3.15})). It is 
also remarkable that the result $\alpha_3=0$ holds independently of any 
assumption about the ``quartic'' parameters $z_1$, $z_2$, and $z_3$.

The second subsystem (\ref{ef08b}) can be viewed (after inserting the unique 
solution of the first subsystem) as an overdetermined system for the two 
unknowns $c_5$, $c_6$.  It is then easily seen that it will admit a solution 
if and only if the parameters $z_1$, $z_2$, and $z_3$ satisfy the following 
linear constraint:
\be
\label{z}
8z_1 + 4z_2 + 3z_3 = 6(4-3\nu)\nu.
\ee
This linear constraint forbids us to consider the simplest ``geodesic'' case 
where $z_1=z_2=z_3=0$. We can, however,  continue to impose the natural 
conditions $z_1=0=z_2$ which simplify very much the 3PN effective dynamics of 
circular orbits. With this choice, the general constraint (\ref{z}) yields
\be
z_3 = 2(4-3\nu)\nu.
\ee
Having fixed the values of $z_1$, $z_2$, and $z_3$, the system (\ref{ef08b}) 
uniquely determines $c_5$ and $c_6$ to be
\begin{mathletters}
\bea
c_5 &=& -\frac{1}{16}(13-16\nu+6\nu^2)\,\nu,
\\[2ex]
c_6 &=& -\frac{1}{48}(115+116\nu-26\nu^2)\,\nu.
\eea
\end{mathletters}

Finally, the subsystem (\ref{ef08c}) gives 3 equations for the remaining
3 unknowns $c_7$, $d_3$, and $a_4$. The unique solution of this system reads:
\begin{mathletters}
\label{ef09b}
\bea
c_7 &=& -\left(\frac{1}{64}\pi^2+\frac{155}{24}\right)\nu + \frac{3}{8}\nu^2
+ \frac{1}{8}\nu^3,
\\[2ex]
d_3 &=& 2(3\nu-26)\,\nu,
\\[2ex]
\label{a4}
a_4 &=& \left(\frac{94}{3}-\frac{41}{32}\pi^2+2\oms\right)\nu.
\eea
\end{mathletters}

Note how simple the structure of the coefficient $a_4$, Eq.\ (\ref{a4}), is.
Indeed, the right-hand-sides of all the equations (\ref{ef08a}), (\ref{ef08b}),
(\ref{ef08c}), were polynomials of the third degree in $\nu$.  Therefore, one
would have a priori expected the 3PN coefficient $a_4$ to have the same
structure: $a_4=a_{41}\nu+a_{42}{\nu}^2+a_{43}{\nu}^3$. It is
remarkable that the coefficients of $\nu^2$ and $\nu^3$ happen to vanish in
$a_4$ (such a simplification does not occur in the 3PN-level coefficients
appearing in the functions $E(x)$, $e(x)$ and $j^2(x)$ considered above).  This
simple structure of $a_4$ can be brought out by defining the following quantity,
\be
\label{omstar}
\oms^* \equiv -\frac{47}{3} + \frac{41}{64}\pi^2 = -9.3439\ldots,
\ee
in terms of which the value of $a_4$ can be written as
\be
\label{a4omstar}
a_4 = 2(\oms -\oms^*)\,\nu.
\ee
The presence of already two cancellations in $a_4$ ($a_{42}=0=a_{43}$)
suggests that the yet undetermined value of $\oms$ might be precisely $\oms^*$,
so that $a_4$, Eq.\ (\ref{a4}), simply vanishes. We shall see below
that this conjecture is indirectly supported by the fact that a numerical value
$\oms\simeq-9$ is selected by the requirement that the various methods 
discussed in this paper agree in their predictions for LSO quantities.
Note also the  remarkable fact that if $\oms=\oms^*$ all the $\pi^2$ terms
cancell in all the 3PN-level coefficients, so that they all become rational
(as were the 2PN ones). [Stated in reverse, if one could a priori prove that all 
the 3PN coefficients are rational, this would support the conjecture that 
$\oms=\oms^*$, though it would be also compatible with having $\oms=\oms^*+$ a 
rational number.]

The coefficient $a_4$ enters the PN expansion of $A(u)\equiv-g_{00}(q')$ (with 
$u\equiv{1/q'}$):
\be
\label{ef10}
A(u) = 1 - 2u + 2\nu u^3 + a_4(\nu) u^4 + {\cal O}(u^5).
\ee
As mentioned above, we improve the behaviour of the PN expansion of $A(u)$ by 
Padeeing it:
$A_{P_n}(u)=P_\ell^k[T_{n+1}[A(u)]]$ with $k+\ell=n+1$. We impose the constraint 
$k>0$ to inject the information that $A(u)$ should qualitatively look like
$A_{\rm Schw}(u)=1-2u$, i.e.\ have a zero at some $u=\frac{1}{2}+{\cal O}(\nu)$. 
As said above, the most robust (for our purpose) 
Pad\'es of $A(u)$ are the ones with $k=1$ 
and $\ell=n$. Finally, we get the following sequence of Pad\'e-improved $A$:
\begin{mathletters}
\label{ef11}
\begin{eqnarray}
A_{P_1}(u) &\equiv& P^1_1\left[T_2[A(u)]\right] = 1-2u\,,
\\[2ex]
A_{P_2}(u) &\equiv& P^1_2\left[T_3[A(u)]\right]
= \frac{1-\left(2-\frac{1}{2}\nu\right)u}{1+\frac{1}{2}\nu u+\nu u^2}\,,
\\[2ex]
A_{P_3}(u) &\equiv& P^1_3\left[T_4[A(u)]\right]
= \frac{2(4-\nu)+\biglb(a_4(\nu)-16+8\nu\bigrb)u}
{2(4-\nu)+\biglb(a_4(\nu)+4\nu\bigrb)u+2\biglb(a_4(\nu)+4\nu\bigrb)u^2
+4\biglb(a_4(\nu)+\nu^2\bigrb)u^3}\,.
\end{eqnarray}
\end{mathletters}

To extract the LSO quantities from these Pad\'eed $A$'s, we must consider the 
effective radial potential
\be
\label{ef12}
W_j^{P_n}(u) = A_{P_n}(u)(1+j^2 u^2)\,.
\ee
The value of $j$ for which this radial potential has an inflection point defines 
$j_{\rm LSO}(\nu)$; the corresponding value of $u$ being
$u_{\rm LSO}=u_*(j_{\rm LSO})$. As explained in Eqs.\ (\ref{eq3.22}) and 
(\ref{eq3.23}) above one then deduces the energy and the orbital frequency of 
the LSO. Note that the 2PN Pad\'eed $A_{P_2}$ that we use here differs from the 
straightforward Taylor approximant $A_{T_2}$ used in Ref.\ \cite{BD99}. However, 
this difference is essentially negligible (as shown by comparing the lines
``eff. method'' and ``BD'' in Table I). The Pad\'e improvement is, however, 
rather important at 3PN in the case where $\oms$ is significantly larger than
$\oms^*$. Indeed, in this case, Pad\'eeing  allows one to tame the effect of a 
largish (positive) 
$a_4(\nu)$: $a_4\left(\frac{1}{4}\right)\simeq4.67+\frac{1}{2}\oms$. For 
instance, when $\nu=\frac{1}{4}$ and $\oms\gtrsim-1.2$ the radial potential 
built from a straight Taylor-approximated $A_{T_4}(u)$ would not give rise to an 
inflection point continuously connected to the test-mass limit. We 
attribute this lack of structural stability to the known bad properties of 
high-order PN expansions, and not to the effective-one-body approach.
The (numerical) results obtained by the effective one-body approach are 
exhibited in Table \ref{tab:lso}.

\section{Discussion}

Before discussing the meaning of the results obtained above, let us state what 
we would a priori expect. First, we recall that the study in Ref.\ \cite{DIS} 
has shown that the sequence of Pad\'e approximants of the invariant function 
$F(v)$, giving the gravitational wave flux in terms of 
$v\equiv(GM\omega/c^3)^{1/3}=x^{1/2}$, had very good (and very monotonic) 
convergence properties toward the exact result. [By contrast, the sequence of 
Taylor approximants was badly convergent, and unstable when 
$v\lesssim v_{\rm LSO}^{\rm Schw}=0.40825$; see Figs.\ 3a and 3b of \cite{DIS}.] 
In our case, as one can meaningfully (at least for the $j$- and 
effective-one-body methods) consider the 1PN, 2PN, and 3PN approximations, we 
would expect that a good resummation technique would ensure that any LSO 
quantity, say $Q_{\rm LSO}$, be determined with increasing accuracy, when using 
higher PN information, and, more precisely, that
\be
\label{eq5.1}
Q_{\rm LSO}^{P_n} \simeq Q_{\rm LSO}^X + a(b\,x_{\rm LSO})^{n+1},
\ee
with (hopefully) coefficients $a$ and $b$ small enough to ensure a visible 
convergence (when $x_{\rm LSO}\simeq x_{\rm LSO}^{\rm Schw}=1/6$). As a 
minimum test of improved convergence we hope that $|Q_{\rm 3PN}-Q_{\rm 2PN}|$ 
would be significantly smaller than $|Q_{\rm 2PN}-Q_{\rm 1PN}|$, i.e.\ that the 
addition of the 3PN information would have only slightly refined the previous 
2PN estimates of LSO quantities \cite{DIS,BD99}.

Independently of this expectation, we had also hoped, when starting this
investigation, that the LSO quantities might be ``robust'' under the lack of
precise knowledge of a sole ambiguous coefficient ($\oms$) among many others in
$H_{\rm 3PN}$.  [Given that the amplitude of this coefficient would have some
plausible upper bound; as discussed in the Appendix A of \cite{DJS1}.]  However,
the results exhibited in Table \ref{tab:lso} and Fig.\ \ref{fig:js} show that,
in spite of our use of resummation techniques, the LSO quantities appear to be
quite sensitive to the exact value of $\oms$.  A first conclusion of our work is
therefore that it is quite important to resolve the problem of static ambiguity,
arising at 3PN when using delta functions to represent compact (but extended)
objects.  Until this problem is unambiguously solved, it will not be clear
whether (as proposed in \cite{DIS,BD99}) it is possible to trust suitably
resummed versions of PN-expanded results.

\begin{figure}
\centerline{\epsfig{file=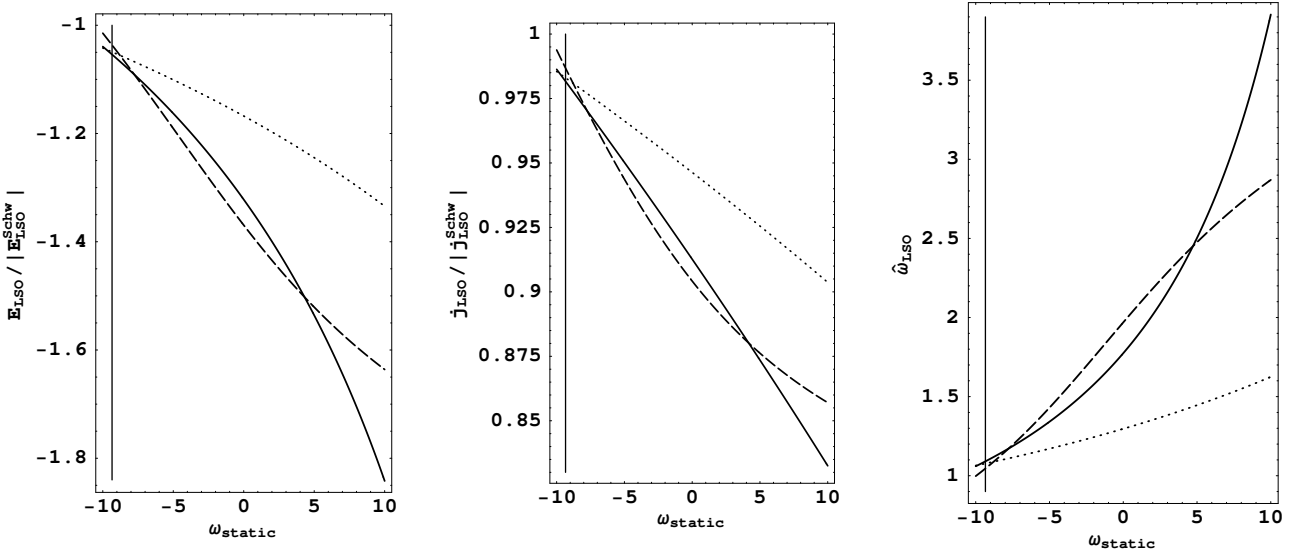,viewport=0 550 612 792}}
\caption{\label{fig:sigma}
Equal-mass ($\nu=\frac{1}{4}$) binary systems LSO parameters obtained by means 
of different methods as functions of the ambiguity parameter $\oms$. The 
reduced binding energy $E_{\rm LSO}$, Eq.\ (\ref{eq2.2}), and the reduced 
angular momentum $j_{\rm LSO}$, Eq.\ (\ref{eq2.17}), are divided by their 
``Schwarzschild values'':
$\left\vert{E_{\rm LSO}^{\rm Schw}}\right\vert=1-\frac{1}{3}\sqrt{8}$ and
$j_{\rm LSO}^{\rm Schw}=\sqrt{12}$; the dimensionless orbital frequency 
$\widehat{\omega}_{\rm LSO}$ is defined in Eq.\ (\ref{e08}). The lines shown in 
the plots correspond to different methods: $j$-method (solid), effective 
one-body method (dotted), and $e$-method (dashed). The solid vertical lines 
correspond to $\oms=\oms^*$, Eq.\ (\ref{omstar}).}
\end{figure}

In the meantime, however, we wish to point out several remarkable features of 
the dependence of our various results on $\oms$. In Fig.\ \ref{fig:sigma} we 
plot (for the equal-mass case, $\nu=1/4$) our various predictions, at the 3PN 
level, and using various methods, as a function of the 3PN ambiguity parameter 
$\oms$. It is quite interesting to note that two a priori independent things 
happen:

(i) there is a value of $\oms$, namely
\be
\label{eq5.2}
\oms^{\text{best}} \simeq -9,
\ee
for which the three different methods give, at 3PN, nearly coincident LSO 
predictions.

(ii) For this ``best fit'' value $\oms^{\text{best}}$ the 3PN LSO 
predictions exhibit the expected convergence property that
$|Q_{\rm 3PN}-Q_{\rm 2PN}|$ is significantly smaller than
$|Q_{\rm 2PN}-Q_{\rm 1PN}|$ (see Table \ref{tab:shanks} below).

We have checked that these two remarkable properties hold for all values of the
parameter $\nu\leq\frac{1}{4}$.  Actually there are several other ways of
selecting the approximate value (\ref{eq5.2}), i.e.\ of understanding why it
plays a special role.  First, we have seen above that the precise value
(\ref{omstar}) (which is near the ``best fit'' value (\ref{eq5.2})), played a
special role in simplifying not only $a_4$ but also all the other 3PN
coefficients.  Second, it seems natural to expect that the true value of $\oms$
will be such that the full Taylor expansions of most of the invariant functions
will be smooth deformations of their test-mass limits.  A minimum requirement
for this property of ``structural stability'' under the turning-on of the
parameter $\nu$ seems to be that the Taylor coefficients of the functions
$e(x)$, $j^2(x)$, $K(y)$, and $1/A(u)$ do not change sign as $\nu$ varies from
0 to 1/4 (we only consider functions with infinitely many non-zero Taylor
coefficients in the test-mass limit).  [One could actually impose a more
restricted bound on the $\nu-$variation of the 3PN coefficients, especially
given the information that the 2PN coefficients are found to vary by a smallish
fractional amount.]  We find that this minimum requirement is satisfied only if
$\oms<-0.62$ (the consideration of the expansion of $e(x)$ gives the most
stringent bound).  Another natural requirement for ``structural stability''
under $\nu \neq 0$ would be to impose that all the near-diagonal Pad\'es
(without restricting oneself, as above, to the most ``robust'' ones) of the
functions above exhibit real poles that are smoothly connected to their
test-mass counterparts.  This requirement is most stringent when considering
$P^2_2(1/A(u))$, and yields the limit $\oms < -8.35$.  Combining this with the
general limits Eq.\ (\ref{oms}) suggests that $\oms$ lies within the small range 
$-10<\oms<-8.35$.

Let us quote a last way of selecting the value (\ref{eq5.2}).  It consists in 
comparing the 3PN Taylor coefficients of the invariant functions that contain 
only a finite number of terms in the test-mass limit.  For instance, consider 
$A(u)$ and $1/j^2(x)$.  In the test-mass limit $A(u)=1-2u$ and 
$1/j^2(x)=x-3x^2$.  When $\nu\neq0$ there will appear further powers of $u$ or 
$x$ with coefficients vanishing with $\nu$.  At the 3PN level there is a term 
$a_4u^4$ in $A(u)$, and a term $\frac{8}{3}b_4x^4$ in $1/j^2(x)$.  [The factor 
8/3 is introduced to have the same (linear) dependence on $\oms$ in $a_4$ and 
$b_4$.]  These two coefficients read
\bea
\label{eq5.6}
a_4\,(\nu) &=& \bigg(\frac{94}{3}-\frac{41}{32}\pi^2\bigg)\nu + 2\,\oms\,\nu
\nonumber\\[2ex]
&\approx& 18.6879\,\nu + 2\,\oms\,\nu
\eea
and
\bea
\label{eq5.13}
b_4(\nu) &=& \bigg(\frac{5269}{192}-\frac{41}{32}\pi^2\bigg)\,\nu
- \frac{61}{32}\,\nu^2 + \frac{1}{216}\,\nu^3 + 2\,\oms\,\nu
\nonumber\\[2ex]
&\approx& 14.7973\,\nu - 1.9063\,\nu^2 + 0.00463\,\nu^3 + 2\,\oms\,\nu.
\eea

Let us first note that the terms $\propto\nu^2$ and $\nu^3$ in Eq.\
(\ref{eq5.13}) are numerically nearly negligible.  Forgetting about them (i.e.\
working with $b'_4\equiv (\frac{5269}{192}-\frac{41}{32}\pi^2)\,\nu$), we then 
see, by comparing Eqs.\ (\ref{eq5.6}) and (\ref{eq5.13}), that the
coefficients $a_4(\nu)$ and $b_4^{\prime}(\nu)$ are approximately identical.  In
particular, this means that there will be a small range of values of $\oms$ for
which $a_4$ and $b_4^{\prime}$ will be {\em simultaneously} small.  The
existence of this range explains why the $j$- and effective one-body methods can
give numerically similar results.  We can then make an {\em analytical} estimate
of the `best' value of the ambiguity parameter $\oms$ by looking for the value
of $\oms$ which {\em simultaneously} minimizes (in a least square sense) $a_4$
and $b'_4$.  It is easily seen that the expression
$[a_4(\nu)]^2+[b'_4(\nu)]^2$ attains its minimal value (as function of $\oms$) 
for
\be
\label{eq5.15}
\oms^{\text{min}} = \frac{41}{64}\pi^2-\frac{11285}{768} \approx -8.37\,.
\ee
This numerical value is not very far the special values selected by the 
other arguments discussed above.

Summarizing: several (partially) independent arguments suggest that the true
value of $\oms$ lies in the range $ -10 \alt \oms \alt -8$. For definiteness,
and for the purpose of the following discussion, we shall henceforth assume
that the ``correct'' value of $\oms$ is
\be
\label{eq5.3}
\oms = \oms^*.
\ee
In Fig.\ \ref{fig:sigma} we have included vertical lines corresponding to 
$\oms=\oms^*$, to show visually that Eq.\ (\ref{eq5.3}) is well compatible with 
our argument based on the convergence of the various methods.

One can also see in Fig.\ \ref{fig:sigma} that the curves related to the $e$-
and $j$-methods have a second intersection point, besides the one around
$\oms^{\text{best}}\simeq-9$.  However, we have checked that the value of $\oms$
at this point strongly depends on the value of the parameter $\nu$.  For this
reason, and also for the fact that this point does not give an agreement with
the ``effective'' method, we do not take this second intersection point as
evidence for a different value of $\oms$.

Admitting (for the sake of the following argument) Eq.\ (\ref{eq5.3}) we wish to
propose a {\it further} way of improving the accuracy of the predictions of LSO
observables.  Indeed, if one has at one's disposal {\it three} successive
approximations, namely the 1PN, 2PN, {\it and} 3PN estimates of some quantity
$Q_{\rm LSO}$, one can combine this information to refine the estimate of the
(unknown) exact value $Q_{\rm LSO}^X$.  The rationale for this is to assume that
the approach to the limit, when the order $n$ of the approximant increases, is
approximately described by Eq.\ (\ref{eq5.1}) (i.e.\ essentially that the
inaccuracy of the $n$th estimate decreases proportionally to the $(n+1)$th power
of some constant $c\equiv b\,x_{\rm LSO}<1$).  Then, under this assumption the
knowledge of three (successive) approximants, say $Q_{n-1}$, $Q_n$, and
$Q_{n+1}$, gives three equations ($Q_m=Q_X+a\,c^{m+1}$) for the three unknowns
$(Q_X,a,c)$.  One can solve this system of equations and deduce, in particular,
the value of the looked for $n\to\infty$ limit $Q_X$ in terms of $Q_{n-1}$,
$Q_n$, and $Q_{n+1}$.  The result defines the so-called ``Shanks
transformation'' \cite{BO84}, namely
\be
\label{eq5.4}
Q_X \simeq S_n\,[Q]
\equiv \frac{Q_{n+1}\,Q_{n-1}-Q_n^2}{Q_{n+1}+Q_{n-1}-2\,Q_n} \,.
\ee
When one disposes of more than three $Q_m$'s, the Shanks transformation 
associates to the original (truncated) sequence $(Q_1,Q_2,\ldots,Q_N)$ a 
shorter, but hopefully faster converging sequence 
$(S_2\,[Q],\ldots,S_{N-1}\,[Q])$. In our case, the Shanks procedure associates 
to any triplet of LSO quantities $(Q_1,Q_2,Q_3)
\equiv(Q_{\rm 1PN}^{\rm LSO},Q_{\rm 2PN}^{\rm LSO},Q_{\rm 3PN}^{\rm LSO})$ a 
single number,
\be
\label{eq5.5}
Q_S^{\rm LSO} \equiv
\frac{Q_{\rm 3PN}^{\rm LSO}\,Q_{\rm 1PN}^{\rm LSO}-(Q_{\rm 2PN}^{\rm LSO})^2}
{Q_{\rm 3PN}^{\rm LSO}+Q_{\rm 1PN}^{\rm LSO}-2\,Q_{\rm 2PN}^{\rm LSO}},
\ee
which is a (hopefully better) estimate of the (unknown) exact value
$Q_{\rm LSO}^X$. We shall refer to (\ref{eq5.5}) as the $S$-estimate of
$Q^{\rm LSO}$.

\begin{figure}
\centerline{\epsfig{file=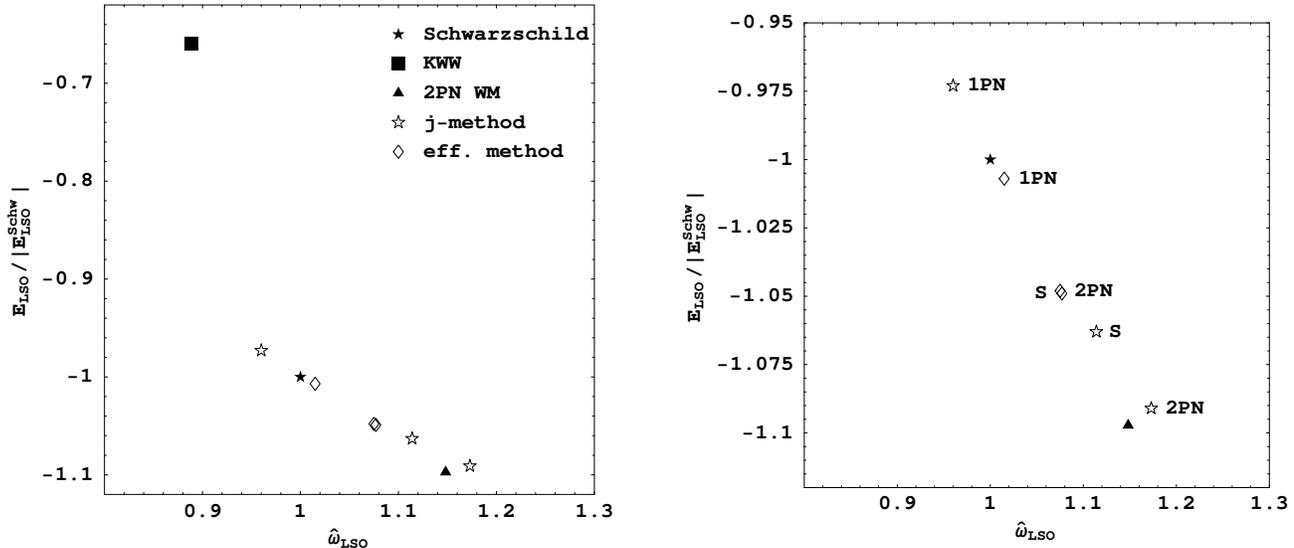,viewport=0 534 612 792}}
\caption{\label{fig:envsom}
Equal-mass ($\nu=1/4$) binary systems reduced binding energy
$E_{\rm LSO}/\vert{E_{\rm LSO}^{\rm Schw}}\vert$ versus the dimensionless 
orbital frequency $\widehat{\omega}_{\rm LSO}$, for different methods discussed 
in our paper. For $j$- and effective one-body methods we have plotted the 
results at the 1PN level, 2PN level, and $S$-approximants; they are all 
exhibited in Table \ref{tab:shanks}. We have also shown the results obtained in 
Refs.\ [8] (labelled by KWW), and by applying the 2PN effective-one-body 
method to the ``Wilson-Mathews'' truncation of general relativity (labelled by 
WM, see Appendix B).}
\end{figure}

In Table \ref{tab:shanks} (see also Fig.\ \ref{fig:envsom}) we apply this 
procedure to our two best methods: the $j$-method and the effective-one-body 
one, under the assumption (\ref{eq5.3}) (which is needed to exhibit a visible 
convergence among the first three PN approximations).

\begin{table}
\caption{Equal-mass ($\nu=1/4$) binary systems LSO parameters obtained by means 
of the $j$- and effective-one-body methods. The 3PN values of the LSO parameters 
were calculated for $\oms=\oms^*$, Eq.\ (\ref{omstar}); Eq.\ (\ref{eq5.5}) 
defines the $S$-estimates.}
\label{tab:shanks}
\begin{center}
\begin{tabular}{c|cccc|cccc|cccc}
& \multicolumn{4}{c|}{$E_{\rm LSO}/\vert{E_{\rm LSO}^{\rm Schw}}\vert$}
& \multicolumn{4}{c|}{$j_{\rm LSO}/j_{\rm LSO}^{\rm Schw}$}
& \multicolumn{4}{c}{$\widehat{\omega}_{\rm LSO}$} \\
method & 1PN & 2PN & 3PN & $S$-estimate
       & 1PN & 2PN & 3PN & $S$-estimate
       & 1PN & 2PN & 3PN & $S$-estimate \\ \hline
$j$-method
& $-0.973$
& $-1.091$
& $-1.054$
& $-1.063$
  & 1.014
  & 0.970
  & 0.982
  & 0.979
& 0.960
& 1.173
& 1.091
& 1.114 \\
eff.\ method
& $-1.007$
& $-1.048$
& $-1.049$
& $-1.049$
  & 1.000
  & 0.983
  & 0.983
  & 0.983
& 1.015
& 1.075
& 1.077
& 1.077 \\
\end{tabular}
\end{center}
\end{table}

Given our present (incomplete) knowledge we consider that the $S$-estimates
exhibited in Table \ref{tab:shanks} represent our best estimates of LSO
observables.  To verify the plausiblity of these estimates one should resolve
the issue of the ambiguous coefficient $\oms$ in the 3PN dynamics.  [In
principle, this can be done by implementing the matching method described in
\cite{LH} and used there at the 2PN level.]  If this resolution approximately
confirms the estimate (\ref{eq5.3}) the $S$-estimates will be confirmed.  If a
very different value of $\oms$ is found, it might still be compatible with a
less evidently convergent PN sequence.  And hopefully, the $S$-estimate
(\ref{eq5.5}) of this new sequence will give an improved 3PN-accurate estimate
of LSO observables.

Under the assumption that the $S$-estimates are accurate, there are several
interesting conclusions that we can draw.  First, we remark that the final
estimates are quite near the 2PN-level predictions of the effective one-body
approach, see Fig.\ \ref{fig:envsom}.  Although this may seem disappointing (an
enormous, not yet completed, 3PN work leading to a confirmation of 2PN
estimates), this would be a scientifically very useful conclusion.  Indeed, this
would (in our minds at least) establish the soundness of the philosophy
advocated in Refs.\ \cite{DIS,BD99} and here, namely that resummation methods
can be meaningfully employed to make {\it analytical} predictions concerning
physics near the Last Stable Orbit.  This would then also give support to the
recent work of Buonanno and Damour \cite{BDplunge} in which the 2PN effective
one-body Hamiltonian has been used, together with Pad\'e-resummed estimates of
gravitational-radiation damping, to study the transition between the inspiral
motion and the final plunge of a binary system.  [Let us note, in passing, that
this work shows that, though it is crucial to have good initial estimates of the
LSO quantities defined by the Hamiltonian, the final observable effects linked
to the presence of an LSO are blurred by radiation-reaction effects.]

\section*{Acknowledgments}

This work was supported in part by the KBN Grant No.\ 2 P03B 094 17 (to P.J.)  
and the Max-Planck-Gesellschaft Grant No.\ 02160-361-TG74 (to G.S.).
P.J.\ and G.S.\ thank the Institut des Hautes \'Etudes Scientifiques for 
hospitality during crucial stages of the collaboration.

\appendix

\section{3PN effective ``geodesic'' one-body dynamics}

We consider here an effective `relativistic' one-body Hamiltonian 
$\widehat{H}_{\text{eff}}^{\rm R}$ of the simple ``geodesic'' form
\be
\label{app01}
\widehat{H}_{\rm eff}^{\rm R}({\mathbf q}' , {\mathbf p}')
= \sqrt{A (q') \left[ 1 + {\mathbf p}'^2 + \left( \frac{A(q')}{D(q')} - 
1 \right) ({\mathbf n}' \cdot {\mathbf p}')^2 \right]} \,.
\ee
The Hamiltonian $\widehat{H}_{\text{eff}}^{\rm R}$ is related to the real 
`non-relativistic' Hamiltonian $\widehat{H}_{\rm real}^{\rm NR}$, Eq.\ 
(\ref{ef04}), through the constraint equation
\be
\label{app02}
\left[\widehat{H}_{\rm eff}^{\rm R}\biglb(q'(q,p'),p'\bigrb)\right]^2
= \left\{ 1 + \widehat{H}_{\rm real}^{\rm NR}\biglb(q,p(q,p')\bigrb)
\left[ 1 + \alpha_1\,\widehat{H}_{\rm real}^{\rm NR}
+ \alpha_2\left(\widehat{H}_{\rm real}^{\rm NR}\right)^2
+ \alpha_3\left(\widehat{H}_{\rm real}^{\rm NR}\right)^3 \right] \right\}^2.
\ee
As in the text both sides of Eq.\ (\ref{app02}) are written in terms of the 
variables $q$ and $p'$ by means of Eqs.\ (\ref{ef01}) with a generating function 
$G$ of the form
\be
\label{app03}
G(q,p') = G_{1{\rm PN}}(q,p') + G_{2{\rm PN}}(q,p') + G_{3{\rm PN}}(q,p'),
\ee
where
\begin{mathletters}
\label{app04}
\bea
G_{1{\rm PN}}(q,p') &=& ({\mathbf q}\cdot{\mathbf p}')
\left( g_1\,{\mathbf p}'^2 + \frac{g_2}{q} \right), 
\\[2ex]
G_{2{\rm PN}}(q,p') &=& ({\mathbf q}\cdot{\mathbf p}')
\left[ b_1\,{\mathbf p}'^4 + \frac{1}{q}
\biglb(b_2\,{\mathbf p}'^2+b_3({\mathbf n}\cdot{\mathbf p}')^2\bigrb)
+ \frac{b_4}{q^4} \right],
\\[2ex]
G_{3{\rm PN}}(q,p') &=& ({\mathbf q}\cdot{\mathbf p}')
\left[ c_1\,{\mathbf p}'^6 + \frac{1}{q}
\biglb( c_2\,{\mathbf p}'^4
+ c_3\,{\mathbf p}'^2 ({\mathbf n}\cdot{\mathbf p}')^2
+ c_4({\mathbf n}\cdot{\mathbf p}')^4 \bigrb)
+ \frac{1}{q^2} \biglb( c_5\,{\mathbf p}'^2
+ c_6({\mathbf n}\cdot{\mathbf p}')^2 \bigrb) + \frac{c_7}{q^3} \right] \,. 
\eea
\end{mathletters}

Written explicitly, the constraint equation (\ref{app02}) is equivalent to a 
sytem of 23 equations for the 23 unknowns: $a_1$, \ldots, $a_4$; $d_1$, $d_2$, 
$d_3$; $\alpha_1$, $\alpha_2$, $\alpha_3$; $g_1$, $g_2$; $b_1$, \ldots, $b_4$; 
$c_1$, \ldots, $c_7$.
The {\em unique} solution to these equations reads
\begin{mathletters}
\label{app06}
\bea
a_1 &=& -2,
\\[2ex]
a_2 &=& \frac{(3\nu-4)\nu}{2(5-2\nu)},
\\[2ex]
a_3 &=& \frac{(1600-1576\nu+392\nu^2-9\nu^3)\nu}{16(5-2\nu)^2},
\\[2ex]
a_4 &=& \frac{(4280-3349\nu+692\nu^2-9\nu^3)\nu}{6(5-2\nu)^2}
- \frac{41\pi^2}{32}\nu + 2\oms\,\nu,
\\[2ex]
d_1 &=& \frac{(3\nu-4)\nu}{2(5-2\nu)},
\\[2ex]
d_2 &=& \frac{(-2400+1936\nu-408\nu^2+9\nu^3)\nu}{16(5-2\nu)^2},
\\[2ex]
d_3 &=& \frac{(-486400+703680\nu-383904\nu^2+93704\nu^3-8580\nu^4-27\nu^5)\nu}
{64(5-2\nu)^3},
\\[2ex]
\alpha_1 &=& \frac{(4\nu-3)\nu}{2(5-2\nu)},
\\[2ex]
\alpha_2 &=& \frac{(80-32\nu+7\nu^2)(3\nu-4)\nu}{4(5-2\nu)^2},
\\[2ex]
\alpha_3 &=& 
\frac{(3650-3660\nu+1829\nu^2-421\nu^3+44\nu^4)(3\nu-4)\nu}{8(5-2\nu)^3},
\\[2ex]
g_1 &=& \frac{(\nu-6)\nu}{4(5-2\nu)},
\\[2ex]
g_2 &=& \frac{5(4-2\nu+\nu^2)}{4(5-2\nu)},
\\[2ex]
b_1 &=& \frac{(210-178\nu+50\nu^2-3\nu^3)\nu}{16(5-2\nu)^2},
\\[2ex]
b_2 &=& \frac{(-350+419\nu-149\nu^2+13\nu^3)\nu}{8(5-2\nu)^2},
\\[2ex]
b_3 &=& \frac{(120-81\nu+38\nu^2-6\nu^3)\nu}{8(5-2\nu)^2},
\\[2ex]
b_4 &=& \frac{200+200\nu-816\nu^2+360\nu^3-49\nu^4}{32(5-2\nu)^2},
\\[2ex]
c_1 &=& \frac{(5900-12700\nu+8270\nu^2-2082\nu^3+192\nu^4-11\nu^5)\nu}
{64(5-2\nu)^3},
\\[2ex]
c_2 &=& \frac{(-25850+44320\nu-24876\nu^2+5157\nu^3-283\nu^4+13\nu^5)\nu}
{32(5-2\nu)^3},
\\[2ex]
c_3 &=& \frac{(-25200+21360\nu-4364\nu^2-1077\nu^3+408\nu^4-47\nu^5)\nu}
{96(5-2\nu)^3},
\\[2ex]
c_4 &=& \frac{(2160-1834\nu+900\nu^2-228\nu^3+23\nu^4)\nu^2}{32(5-2\nu)^3},
\\[2ex]
c_5 &=& \frac{(247000-407080\nu+225416\nu^2-49240\nu^3+3730\nu^4-151\nu^5)\nu}
{128(5-2\nu)^3},
\\[2ex]
c_6 &=& \frac{7(10300-16360\nu+7136\nu^2-1372\nu^3+154\nu^4-7\nu^5)\nu}
{192(5-2\nu)^3},
\\[2ex]
c_7 &=& \frac{(-962800+1472880\nu-813024\nu^2+206500\nu^3
-25110\nu^4+1479\nu^5)\nu}{384(5-2\nu)^3} - \frac{\pi^2}{64}\nu.
\eea
\end{mathletters}
As said in the text, in view of the complexity of these results, we do not take 
this possibility seriously. We prefer to it the non-minimal (``non-geodesic'')
Hamiltonian given in Sec.\ IV.

\section{2PN results for the conformally-flat truncation of general relativity}

By contrast, let us note that other approximation philosophies are, in our
opinion, less reliable to make predictions concerning the LSO.  We have in mind
here: (i) the use of non-resummed (or only partially resummed) PN expansions,
and (ii) the ``Wilson-Mathews''-type \cite{WM95} truncation of Einstein's
theory, in which the spatial metric is taken to be conformally flat.  As an
example of the first philosophy, let us consider the proposal of Kidder, Will,
and Wiseman \cite{KWW} to partially resum the Damour-Deruelle equations of
motion by separating out (and resumming) the ``Schwarzschild'' $(\nu=0)$ terms.
This approach led to the prediction (at 2PN) that the LSO is significantly less
bound (when $\nu=1/4$) than the ``Schwarzschild'' limit.  In terms of orbital
frequency at the LSO, Ref.\ \cite{KWW} predicts
$\widehat{\omega}_{\rm LSO}(1/4)\simeq0.891<1$.
This contrasts very much with our 2PN
and 3PN estimates above which consistently indicate that $\widehat{\omega}_{\rm
LSO}(\nu)$ is larger than one (and that the LSO is more bound than its
Schwarzschild limit:  $E_{\rm LSO}/\vert{E_{\rm LSO}^{\rm Schw}}\vert<-1$).
Independently of this (biassed) argument, we think that both the robustness and
the consistency of the ``hybrid'' approach of \cite{KWW} are seriously in doubt.
Indeed, Refs.\ \cite{WS93} and \cite{SW93} have shown that the hybrid approach
was robust neither under a change of formulation (Hamiltonian versus
equations-of-motion), nor under a change of coordinate system.  Moreover, Ref.\
\cite{DIS} has questioned the consistency of this approach by pointing out that
the non-resummed ``$\nu$-corrections'' represent, in several cases, a very large
(larger than 100\%) modification of the corresponding $\nu$-independent terms.

Regarding the conformally-flat truncation it was noted by Rieth \cite{R97} that 
this implies significant deviations from the Einstein dynamics already at the 
2PN level. We have investigated this question further. In Ref.\ \cite{DJS1} we 
gave the invariant functions defined (at 2PN accuracy) by the 
Wilson-Mathews-type truncation. Applying now our $j$-method, we find (at 2PN)
\be
\label{eq5.7}
j_{\rm WM}^2(x) = \frac{1}{x} \left[ 1  + \frac{1}{3}\left(9+\nu\right)x
+ \frac{1}{36}\left(324-333\nu-50\nu^2\right)x^2 \right],
\ee
whose Padeed form is
\be
\label{eq5.8}
j_{{\rm WM}P_2}^2(x) \equiv P_1^1\left[T_2[j_{\rm WM}^2(x)]\right]
= \frac{1+\frac{1}{9}\nu+\left(\frac{15}{4}\nu+\frac{1}{2}\nu^2\right)x}
{x \biglb(1+\frac{1}{9}\nu-\left(3-\frac{37}{12}\nu-\frac{25}{54}\nu^2\right)x
\bigrb)}.
\ee
This leads to a prediction for the 2PN $x_{\rm LSO}$ which can be written down 
analytically
\be
\label{eq5.9}
6\,x_{\rm LSO}^{j_{{\rm WM}P_2}}(\nu) = \frac{8(9+\nu)}{3(15+2\nu)\nu}
\left[\frac{2(9+\nu)}{\sqrt{324-333\nu-50\nu^2}}-1\right].
\ee
The corresponding dimensionless orbital frequency, for $\nu=1/4$, equals 
$\widehat{\omega}_{\rm LSO}=1.4378$, the reduced binding energy
$E_{\rm LSO}/\vert{E_{\rm LSO}^{\rm Schw}}\vert=-1.2253$, and the reduced 
angular momentum $j_{\rm LSO}/j_{\rm LSO}^{\rm Schw}=0.9293$.

We have also studied, at the 2PN level, the effective one-body method for the 
Wilson-Mathews dynamics. Using the procedure decribed in Sec.\ IV~D above, 
imposing at the 1PN level the condition $d_1=0$, we have found that the 
effective-metric function $A_{\rm WM}(u)$ at the 2PN accuracy reads
\be
\label{eq5.10}
A_{\rm WM}(u) = 1 - 2u + a_3(\nu) u^3 + {\cal O}(u^4),
\ee
where
\be
\label{eq5.11}
a_3(\nu) = \frac{1}{4}(18-5\nu)\nu.
\ee
We have improved the behaviour of the 2PN expansion of $A_{\rm WM}(u)$ by 
Padeeing it:
\be
\label{eq5.12}
A_{{\rm WM}P_2}(u) \equiv P^1_2\left[T_3[A_{\rm WM}(u)]\right]
= \frac{1-\left(2-\frac{9}{8}\nu+\frac{5}{16}\nu^2\right)u}
{1+\left(\frac{9}{8}\nu-\frac{5}{16}\nu^2\right)u
+\left(\frac{9}{4}\nu-\frac{5}{8}\nu^2\right)u^2}\,.
\ee
To extract the LSO quantities from this Pad\'eed $A$, we have considered the 
inflection point of the effective radial potential
$A_{{\rm WM}P_2}(u)(1+j^2u^2)$, which defines the angular momentum
$j_{\rm LSO}(\nu)$ and the location $u_{\rm LSO}(\nu)$ of the LSO; then using 
Eqs.\ (\ref{eq3.22}) and (\ref{eq3.23}) one calculates the energy and the 
orbital frequency of the LSO. The results, for $\nu=1/4$, are: dimensionless 
orbital frequency $\widehat{\omega}_{\rm LSO}=1.1482$,  reduced binding 
energy $E_{\rm LSO}/\vert{E_{\rm LSO}^{\rm Schw}}\vert=-1.0972$, and  reduced 
angular momentum $j_{\rm LSO}/j_{\rm LSO}^{\rm Schw}=0.9647$.
Let us also mention that Cook, using another conformally flat approximation 
\cite{C94}, obtained, for $\nu=1/4$, the following LSO parameters: the 
dimensionless orbital frequency $\widehat{\omega}_{\rm LSO}=2.528$, the reduced 
binding energy $E_{\rm LSO}/\vert{E_{\rm LSO}^{\rm Schw}}\vert=-1.579$, and the 
reduced angular momentum $j_{\rm LSO}/j_{\rm LSO}^{\rm Schw}=0.8591$.
Similar results have been obtained in Ref.\ \cite{B00}.


\begin{references}

\bibitem{DIS}
T.\ Damour, B.\ R.\ Iyer, and B.\ S.\ Sathyaprakash,
Phys.\ Rev.\ D {\bf 57}, 885 (1998).

\bibitem{DIS2}
T.\ Damour, B.\ R.\ Iyer, and B.\ S.\ Sathyaprakash,
Phys.\ Rev.\ D (accepted), gr-qc/0001023.

\bibitem{CE77}
J.\ P.\ A.\ Clark and D.\ M.\ Eardley,
Astrophys.\ J.\ {\bf 215}, 311 (1977).

\bibitem{BD92}
J.\ K.\ Blackburn and S.\ Detweiler,
Phys.\ Rev.\ D {\bf 46}, 2318 (1992).

\bibitem{C94}
G.\ B.\ Cook,
Phys.\ Rev.\ D {\bf 50}, 5025 (1994).

\bibitem{B00}
T.\ W.\ Baumgarte,
gr-qc/0004050.

\bibitem{DD}
T.\ Damour and N.\ Deruelle,
Phys.\ Lett.\ A {\bf 87}, 81 (1981);
T.\ Damour,
C.\ R.\ S\'eances Acad.\ Sci., S\'er.\ 2 {\bf 294}, 1355 (1982).

\bibitem{DS85}
T.\ Damour and G.\ Sch\"afer,
Gen.\ Rel.\ Grav. {\bf 17}, 879 (1985).

\bibitem{DS88}
T.\ Damour and G.\ Sch\"afer,
Nuovo Cimento B {\bf 101}, 127 (1988).

\bibitem{KWW}
L.\ E.\ Kidder, C.\ M.\ Will, and A.\ G.\ Wiseman,
Class.\ and Quantum Grav.\ {\bf 9}, L127 (1992);
Phys.\ Rev.\ D {\bf 47}, 3281 (1993).

\bibitem{WS93}
N.\ Wex and G.\ Sch\"afer,
Class.\ and Quantum Grav.\ {\bf 10}, 2729 (1993).

\bibitem{SW93}
G.\ Sch\"afer and N.\ Wex,
Phys.\ Lett.\ A {\bf 174}, 196 (1993);
{\bf 177}, 461(E) (1993).

\bibitem{BD99}
A.\ Buonanno and T.\ Damour,
Phys.\ Rev.\ D {\bf 59}, 084006 (1999).

\bibitem{JS98}
P.\ Jaranowski and G.\ Sch\"afer,
Phys.\ Rev.\ D {\bf 57}, 7274 (1998).

\bibitem{DJS2}
T.\ Damour, P.\ Jaranowski, and G.\ Sch\"afer,
Phys.\ Rev.\ D {\bf 61}, 1215XX(R) (2000), gr-qc/0003051.

\bibitem{DJS1}
T.\ Damour, P.\ Jaranowski, and G.\ Sch\"afer,
Phys.\ Rev.\ D {\bf 61}, 1240XX (2000), gr-qc/9912092.

\bibitem{TTS96}
T.\ Tanaka, H.\ Tagoshi, and M.\ Sasaki,
Prog.\ Theor.\ Phys.\ {\bf 96}, 1087 (1996).

\bibitem{JS91}
W.\ Junker and G.\ Sch\"afer,
Mon.\ Not.\ R.\ Astron.\ Soc.\ {\bf 254}, 146 (1992).

\bibitem{BIZ70}
E.\ Br\'ezin, C.\ Itzykson, and J.\ Zinn-Justin,
Phys.\ Rev.\ D {\bf 1}, 2349 (1970).

\bibitem{JS99} 
P.\ Jaranowski and G.\ Sch\"afer,
Phys.\ Rev.\ D {\bf 60}, 124003 (1999).

\bibitem{BF00}
L.\ Blanchet and G.\ Faye, gr-qc/0004009.

\bibitem{BO84}
C.\ M.\ Bender and S.\ A.\ Orszag,
{\it Advanced Mathematical Methods for Scientists and Engineers}
(McGraw-Hill, Singapore, 1984). 

\bibitem{LH}
T.\ Damour,
in {\em Gravitational Radiation},
edited by N.\ Deruelle and T.\ Piran
(North-Holland, Amsterdam, 1983), pp.\ 59--144.

\bibitem{BDplunge}
A.\ Buonanno and T.\ Damour,
Phys.\ Rev.\ D (accepted), gr-qc/0001013.

\bibitem{WM95}
J.\ R.\ Wilson and G.\ J.\ Mathews,
Phys.\ Rev.\ Lett.\ {\bf 75}, 4161 (1995);
J.\ R.\ Wilson, G.\ J.\ Mathews, and P.\ Marronetti,
Phys.\ Rev.\ D {\bf 54}, 1317 (1996).

\bibitem{R97}
R.\ Rieth,
in {\em Mathematics of Gravitation. Part II. Gravitational Wave Detection}, 
edited by A.\ Kr\'olak,
Banach Center Publications
(Institute of Mathematics, Polish Academy of Sciences, Warszawa, 1997),
Vol.\ 41, Part II, pp.\ 71--74.

\end{references}
\end{document}